\documentclass[draftcls, onecolumn, 12pt]{IEEEtran}

\usepackage[english]{babel}
\usepackage[utf8]{inputenc}
\usepackage[T1]{fontenc}
\usepackage[english]{babel}
\usepackage{cite}
\usepackage{color}
\usepackage[cmex10]{amsmath}
\usepackage{amssymb}
\usepackage{amsthm}
\usepackage{bm}
\usepackage{array}
\usepackage{multirow}
\usepackage{colortbl}
\usepackage{graphicx}
\usepackage{epstopdf}
\usepackage[caption=false]{subfig}
\usepackage{booktabs}
\usepackage{hyperref}
\usepackage[nolist]{acronym}

\makeatletter
\hypersetup{ colorlinks     = true,       		   
			 linkcolor      = black,          	   
			 citecolor      = black,               
			 urlcolor       = black,               
			 bookmarksdepth = 4 }
\makeatother

\usepackage{tikz,pgfplots}
\usetikzlibrary{positioning,shapes,arrows,calc,fit,backgrounds,decorations.pathreplacing}
\pgfplotsset{compat=1.12}
\definecolor{mypurple}{RGB}{102,0,204}
\definecolor{mygreen}{RGB}{0,153,0}
\definecolor{myorange}{RGB}{255,128,0}
\definecolor{mybrown}{RGB}{128,64,0}
\tikzstyle{egf}      = [ black, mark=triangle*, mark options={solid,scale=.75}]
\tikzstyle{martin}   = [ red,      mark=diamond*,  mark options={solid,scale=.75}]
\tikzstyle{ofdp}     = [ blue,     mark=*,         mark options={solid,scale=.75}]
\tikzstyle{qcqp1}    = [ mypurple, mark=star,      mark options={solid,scale=.75}]
\tikzstyle{qcqp2}    = [ mygreen,  mark=+,         mark options={solid,scale=.75}]
\tikzstyle{qcqp3}    = [ mybrown,  mark=x,         mark options={solid,scale=.75}]
\pgfplotsset{ legend style = { fill = blue!10,
                 			   legend cell align={left},
                 			   column sep=0.0cm,
                 			   font=\scriptsize,},
			  axisstyle/.style = { width  = .47\textwidth,
					 			   tick label style={/pgf/number format/fixed},
			  		 			   axis background/.style={fill=yellow!10},			
              		 			   grid = major,} }
\pgfplotsset{major grid style={densely dotted,black!30}}
\pgfplotsset{minor grid style={ dotted,black!30}}
\pgfplotsset{every tick label/.append style={font=\scriptsize}}
\pgfplotsset{every label style/.append style={font=\footnotesize}}
\pgfplotsset{legend image code/.code={\draw[mark repeat=2,mark phase=2] plot coordinates {(0cm,0cm) (0.2cm,0cm) (0.4cm,0cm)};}}

\definecolor{gold}{rgb}{0.85,.66,0}

\newcommand{\colr}{\textcolor{red}}

\newcommand{\eye}{\mathbf{I}}
\newcommand{\norm}[1]{\left\Vert {#1} \right\Vert}
\newcommand{\abs}[1]{\left\vert {#1} \right\vert}
\newcommand{\expect}[1]{\mathbb{E}\left[ {#1} \right]}
\newcommand{\real}[1]{\text{Re}\left\lbrace {#1} \right\rbrace}
\newcommand{\imag}[1]{\text{Im}\left\lbrace {#1} \right\rbrace}
\DeclareMathOperator\sinc{sinc}
\newcommand{\mb}[1]{\mathbf{{#1}}}
\newcommand{\rbrac}[1]{\left( {#1} \right)}
\newcommand{\sbrac}[1]{\left[ {#1} \right]}
\newcommand{\cbrac}[1]{\left\lbrace {#1} \right\rbrace}
\newcommand{\ceil}[1]{\left\lceil {#1} \right\rceil}
\newcommand{\floor}[1]{\left\lfloor {#1} \right\rfloor}

\begin{acronym}[\hspace*{3.3cm}] 
\acro{4G}{4th Telecommunication Generation}
\acro{5G}{5th Telecommunication Generation}
\acro{AWGN}{Additive White Gaussian Noise}
\acro{BER}{Bit Error Rate}
\acro{CP}{Cyclic-Prefix}
\acro{DPSS}{Discrete Prolate Spheroidal Sequences}
\acro{DTFT}{Discrete-Time Fourier Transform}
\acro{EGF}{Extended Gaussian Function}
\acro{FBMC}{Filter Bank MultiCarrier}
\acro{GFDM}{Generalized Frequency-Division Multiplexing}
\acro{IAM}{Interference Approximation Method}
\acro{IB}{In-Band}
\acro{ICI}{InterCarrier Interference}
\acro{IOTA}{Isotropic Orthogonal Transform Algorithm}
\acro{ISI}{InterSymbol Interference}
\acro{IoT}{Internet of Things}
\acro{KKT}{Karush-Kuhn-Tucker}
\acro{M2M}{Machine-to-Machine}
\acro{MIMO-FBMC}{Multiple-Input Multiple-Output Filter Bank MultiCarrier}
\acro{MIMO}{Multiple-Input Multiple-Output}
\acro{MMSE-DFE}{Minimum Mean-Square Error Decision Feedback Equalizer}
\acro{MMSE}{Minimum Mean-Square Error}
\acro{MSL}{Maximum Sidelobe Level}
\acro{OFDM}{Orthogonal Frequency-Division Multiplexing}
\acro{OFDP}{Optimal Finite Duration Pulses}
\acro{OoB}{Out-of-Band}
\acro{PAM}{Pulse Amplitude Modulation}
\acro{PAPR}{Peak-to-Average Power Ratio}
\acro{POP}{Pair of Pilots}
\acro{PSK}{Phase-shift keying}
\acro{QAM}{Quadrature Amplitude Modulation}
\acro{QCQP}{Quadratically Constrained Quadratic Programming}
\acro{RMS}{Root Mean Square}
\acro{SDP}{SemiDefinite Programming}
\acro{SIR}{Signal-to-Interference Ratio}
\acro{SISO}{Single-Input Single-Output}
\acro{SNR}{Signal-to-Noise Ratio}
\acro{SOCP}{Second-Order Cone Program}
\acro{SRRC}{Square-Root Raised Cosine}
\acro{UFMC}{Universal Filtered MultiCarrier}
\acro{V-BLAST}{Vertical Bell Labs Layered Space-Time}
\acro{Wimax}{Worldwide Interoperability for Microwave Access}
\acro{ZF}{Zero-Forcing}
\end{acronym}

\begin{document}
\title{FBMC Prototype Filter Design via Convex Optimization}
\author{Ricardo Tadashi Kobayashi and Taufik Abrão
\thanks{
R. T. Kobayashi and T. Abrão are with the Department of  Electrical Engineering (DEEL), State University of Londrina  (UEL), Londrina, PR, Brazil (email: ricardokobayashi.9107@gmail.com, taufik@uel.br).
}
}
\maketitle

\begin{abstract}
In this work, we propose a prototype filter design for \ac{FBMC} systems based on convex optimization, aiming superior spectrum features while maintaining a high symbol reconstruction quality. Initially, the proposed design is written as a non-convex \ac{QCQP}, which is relaxed into a convex \ac{QCQP} guided by a line search. Through the resulting problem, we design three prototype filters: Type-I, II and III. In particular, the Type-II filter shows a slightly better performance than the classical Mirabasi-Martin design, while Type-I and III filters offer a much more contained spectrum than most of the prototype filters suitable for \ac{FBMC} applications. Furthermore, numerical results corroborate the effectiveness of the designed filters as the proposed filters offer fast decay and contained spectrum while not jeopardizing symbol reconstruction in practice.
\end{abstract}
\smallskip
\noindent 

\section{Introduction}\label{sec:fbmc_intro}
Thanks to its simple, yet elegant, implementation and ability to deal with highly selective channels, \ac{OFDM} has become the standard waveform for many contemporary telecommunication systems, including 802.11 for local area networks, 802.16 for Wimax and 4G LTE systems \cite{Hwang2009}. However, \ac{OFDM} presents drawbacks that may be unbearable for 5G and future wireless systems. First, \ac{OFDM} {relies on} \ac{CP} to simplify equalization, which reduces the spectral efficiency. Furthermore, \ac{OFDM} signal deploys a rectangular envelope, leading to a spectrum with high sidelobes (-13 dB), comprising its neighboring bands. Finally, \ac{OFDM} relies strictly on its orthogonality, hence random access channel and multi-cell scenarios may prevent its proper operation.

More recently, an outstanding effort has been spent towards the research on 5G {communication} systems, which envisions applications such as Tactile Internet, Internet of Things and gigabit connectivity \cite{Freeman2016}. Generally speaking, these applications require scalability, robustness, flexibility, low latency, and {improved} spectral/energy efficiencies, which will not be attained exclusively through \ac{OFDM}. Therefore, a paradigm shift on radio access is required to comply the stringent requirements of 5G. In this context, some waveform alternatives have been proposed as 5G candidates, to name a few: \ac{FBMC}, \ac{UFMC}, Filtered \ac{OFDM} and \ac{GFDM} \cite{Wunder2014}.

In special, \ac{FBMC} waveform is regarded as a strong candidate for 5G and other wireless systems to come. By using non-rectangular pulse shaping, \ac{FBMC} generates a lower \ac{OoB} energy emission. Another interesting feature is the ability of \ac{FBMC} to deal with \ac{ISI} without relying on \ac{CP}, making it more efficient than \ac{OFDM}. Also, \ac{FBMC} requirements on time/frequency synchronization are more relaxed than other multicarrier schemes \cite{Cassiau2013}. Yet another important aspect of \ac{FBMC} is its compatibility with massive-MIMO systems \cite{Farhang2014}: one of the main technologies to drive 5G \cite{Boccardi2014}. Indeed, given such promising aspects, \ac{FBMC} waveform was selected as a candidate waveform on METIS project \cite{Metis} and as the main choice for PHYDIAS project \cite{Phydias}. Notice, however, that providing a full and detailed comparison between \ac{FBMC} and \ac{OFDM} is {out} of the scope of this work. In fact, works such as \cite{Fang2013,Waldhauser2006,Farhang2011,Farhang2014} offer a rich comparison between \ac{FBMC} and \ac{OFDM} schemes.

Despite offering many advantages, \ac{FBMC} still presents some issues to be addressed. Due to its dependence on orthogonality in the real field, many algorithms used in other multicarrier systems need to be adapted for \ac{FBMC}. As a result, traditional pilot aided channel estimation cannot be proceed as in \ac{OFDM}, since pilots are prone to imaginary interference at the receiver \cite{Lele2008b,Cui2016}. Furthermore, PAPR reduction techniques need to be adapted for \ac{FBMC}, 
 {e.g.}, Tone Reservation method \cite{Rahmatallah2013,Bulusu2015}.

Another concern that emerges while designing an \ac{FBMC} system is the choice of the prototype filter. For example, when adapting themselves for opportunistic spectrum sharing, cognitive radio{s} should minimize \ac{OoB} emission in order to avoid interfering {with} other bands \cite{Kumar2016}. Although being less vulnerable to time/frequency channel dispersion, \ac{FBMC} signals are still prone to such effects \cite{Zhang2017}. Thus, well localized filters in time and frequency are desirable. Another aspect to be taken into account is the reconstruction capabilities offered by the filter, as \ac{FBMC} systems operate under an intrinsic interference floor dictated by the prototype filter.

From this perspective, this work proposes a prototype filter design based on convex optimization which aims for the power minimization of the \ac{OoB} pulse energy, constrained to a maximum tolerable self-interference level and a fast spectrum decay. Initially, the design is modeled as an optimization problem in the standard form, which is, unfortunately, a non-convex \ac{QCQP}. To circumvent such an issue, we propose a relaxation which leads to a convex \ac{QCQP} guided by a line search. In the sequel, we test our methodology by designing three prototype filters. Numerical results show that the three proposed filters can provide both {proper}  symbol reconstruction and a very desirable spectral response, making them {an exciting} option to be deployed in \ac{FBMC} systems to come.

The contribution of this work is fourfold:
\begin{itemize}
\item A detailed formulation of the optimization problem prototype filter;
\item The proposition of a relaxation that leads to a convex problem;
\item {Design methodology of} three filters with superior spectrum features and symbol reconstruction capabilities compatible with the operation of \ac{FBMC} systems;
\item A fair comparison between the proposed prototype filters and other popular choices.
\end{itemize}

The remainder of the paper is organized as follows. In Section \ref{sec:fbmc_mux}, a brief description on \ac{FBMC} systems is provided. In order to establish a methodology for comparing different prototype filters, Section \ref{sec:fbmc_merit} presents some figures of merit to evaluate the performance provided by prototype filters from different perspectives. Section \ref{sec:proto} makes a quick survey on popular prototype filters in \ac{FBMC} systems, which are used for comparison along the numerical results. In Section \ref{sec:fbmc_proposed_design}, the proposed prototype filter design is presented and discussed. Numerical results are presented in Section \ref{sec:numerical} and Section \ref{sec:conclusions} offers the conclusions and final remarks of the work.

\textbf{Notation:} Vectors are denoted by lower case bold letters and matrices are denoted by upper case bold letters.  $\mb{A}^{-1}$, $\mb{A}^{T}$ and $\norm{\mb{A}}_p$ are the inverse, the transpose and the p-norm of $\mb{A}$, respectively. Also, $\eye$ denotes the identity matrix and $\sbrac{\mb{A}}_{i,k}$ is the entry {for} the $i$th row and $k$th column of $\mb{A}$. Furthermore, $\rho(\mb{A})$ is the set of eigenvalues of $\mb{A}$, $\rho_{\max}(\mb{A})$ is the largest eigenvalue of $\mb{A}$ and $\rho_{\min}(\mb{A})$ is the smallest eigenvalue of $\mb{A}$. $\real{z}$ and $\imag{z}$ are the real part and imaginary part of $z$, receptively, and $j=\sqrt{-1}$. $\ceil{x}$ smallest integer greater than or equal to $x$ and $\floor{x}$ the largest integer that is less than or equal to $x$. Finally, $\expect{\cdot}$ is the expectation operator and $\left< a[k] \left\vert b[k] \right.\right>$ is the inner product of $a[k]$ and $b[k]$.

\section{\ac{FBMC} Multiplexing}\label{sec:fbmc_mux}
An \ac{FBMC} signal consists of a set of staggered {\ac{PAM}} symbols multiplexed along $M$ subcarriers using a particular filtering, which enables features like low \ac{OoB} emission. Differently from \ac{OFDM}, the \ac{FBMC} symbol interval is shorter than its duration, which compensates the amount of data multiplexed, as two staggered PAM symbols are transmitted instead of one {\ac{QAM}} or {\ac{PSK}} symbol. In order to ensure a better orthogonality between different symbols, a suitable phase shift is introduced {into} each PAM symbol and a well designed filter is also required.

At the critical sampling rate, the multiplexed \ac{FBMC} signal can be expressed as \cite[eq. (1)]{Cui2016}
\begin{equation}
	s[k]  =  \displaystyle\sum_{n=-\infty}^{\infty} 
			 \displaystyle\sum_{m=0}^{M-1}
			   a_{m,n}  p_{m,n}[k],
	\label{eq:fbmc_siso_disc}
\end{equation}
where $a_{m,n}$ is the $n$th PAM symbol of the $m$th subcarrier and $p_{m,n}[k]$ is the pulse shape used by such a symbol. In particular,
\begin{equation}
	p_{m,n}[k] = p\sbrac{k-n\dfrac{M}{2}}
			  e^{ j\rbrac{\frac{2\pi}{M} m \underline{k} + \phi_{m,n}}},
\end{equation}
where $p[k]$ is the prototype filter, $\underline{k}=k-(L_p-1)/2$ and $\phi_{m,n}$ is the phase shift introduced into $a_{m,n}$. The prototype filter is truncated to $L_p$ samples and typical values are $KM-1$, $KM$ and $KM+1$, where $K$ is referred as the overlapping factor. Noticeably, the distance between adjacent subcarriers is $1/M$, while the signaling interval is $M/2$, \textit{i.e.}, half the amount of samples used in \ac{OFDM}. Hence, an \ac{FBMC} symbol introduces interference to its neighborhood, which must be addressed in order to enable the symbol reconstruction at the receiver. In this sense, the phase shift $e^{j\phi_{m,n}}$ is introduced to minimize the interference between adjacent symbols. The most common choice used to phase-shift PAM symbols \cite{Phydias,Siohan2002} is
\begin{equation}
	e^{j\phi_{m,n}} = e^{j\frac{\pi}{2}(m+n)},
	\label{eq:fbmc_phase1}
\end{equation}
which makes adjacent symbols to be phase-shifted by $\pi/2$ and will be used throughout this work.

\subsection{Symbol Reconstruction}
Since PAM symbols are deployed to convey information, one may retrieve the symbol $a_{m_0,n_0}$ by taking the real part of the projection of $p_{m_0,n_0}[k]$ onto the multiplexed signal $s[k]$, \textit{i.e.}, 
\begin{equation}
	\tilde{a}_{m_0,n_0} =
	\real{{\left< s[k] \left\vert p_{m_0,n_0}[k] \right.\right>} }.
	\label{eq:fbmc_est1}
\end{equation}
Unfortunately, the set of sequences $\cbrac{p_{m,n}[k]}$ is not orthogonal, even considering an appropriate phase-shift and the real part operator. Differently from \ac{OFDM}, adjacent \ac{FBMC} symbols overlap as the filter length is larger than the signaling interval, \textit{i.e.}, $L_p>M/2$. Thus, the estimated symbol is composed by the symbol itself and its associated interference:
\begin{equation}
	\hspace*{-1mm}
	\tilde{a}_{m_0,n_0} =
	a_{m_0,n_0} +
	\displaystyle\sum_{\substack{ n\neq n_0 \\m\neq m_0}}
	\hspace*{-1mm}
	a_{m,n} 
	\real{\left< p_{m,n}[k] \left\vert p_{m_0,n_0}[k] \right.\right>}. 
	\label{eq:fbmc_est2}
\end{equation}
Hence, an appropriate prototype filter should be employed in order to enable reconstruction at the receiver, since it introduces a self-interference according to eq. \eqref{eq:fbmc_est2}.

\subsection{\ac{FBMC} Transmultiplexer Scheme}
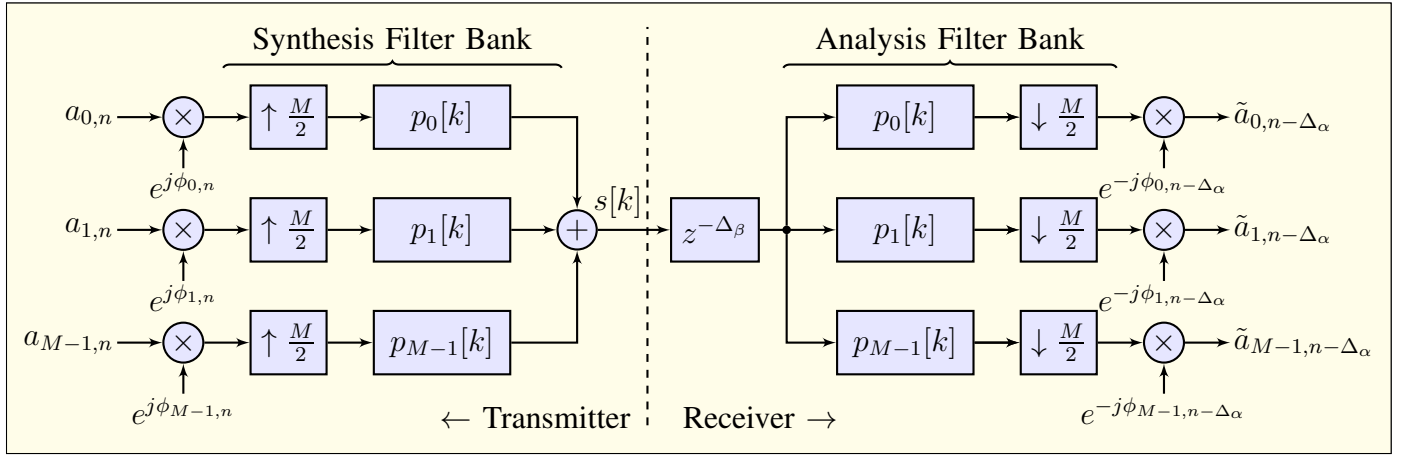
\begin{figure*}[pt]
\begin{center}
\begin{tikzpicture}[ node distance= 1.5cm and .8cm,
                     >=latex',thick,
                     framed,
                     background rectangle/.style={fill=yellow!10,draw=black},
                     show background rectangle,]
	\tikzstyle{circ1} = [draw, circle,inner sep=1pt,fill=blue!10]
	\tikzstyle{rect0} = [minimum width=1em,inner sep=1pt]
	\tikzstyle{rect1} = [draw, rectangle, minimum width=2em, minimum height = 2em,fill=blue!10]
	\tikzstyle{rect2} = [draw, rectangle, minimum width=4.3em, minimum height = 2em,fill=blue!10]
	\tikzstyle{rect3} = [draw, rectangle, minimum width=4.3em, minimum height = 2em,fill=blue!10]
	\tikzstyle{dot1}  = [inner sep=.85pt,fill,draw,circle]
	\def\dx{0.6}
	\def\dya{0.4}
	\def\dyb{1.5}
	\node[rect0,anchor=east]              (in0)  at (0,0)             {$a_{0,n}$};
	\node[circ1,anchor=west]  (mul0) at ($(in0.east)  +(+\dx,+0.0)$)  {$\times$};
	\node[rect0,anchor=north] (phi0) at ($(mul0.south)+(+0.0,-\dya)$)  {$e^{j\phi_{0,n}}$};
	\node[rect1,anchor=west]  (exp0) at ($(mul0.east) +(+\dx,+0.0)$)  {$\uparrow\frac{M}{2}$};
	\node[rect2,anchor=west]  (f0)   at ($(exp0.east) +(+\dx,+0.0)$)  {$p_0[k]$};
	\node[rect0,anchor=east]              (in1)  at ($(0,-\dyb)$)            {$a_{1,n}$};
	\node[circ1,anchor=west]  (mul1) at ($(in1.east)  +(+\dx,+0.0)$)  {$\times$};
	\node[rect0,anchor=north] (phi1) at ($(mul1.south)+(+0.0,-\dya)$)  {$e^{j\phi_{1,n}}$};
	\node[rect1,anchor=west]  (exp1) at ($(mul1.east) +(+\dx,+0.0)$)  {$\uparrow\frac{M}{2}$};
	\node[rect2,anchor=west]  (f1)   at ($(exp1.east) +(+\dx,+0.0)$)  {$p_1[k]$};
	\node[rect0,anchor=east]              (in2)  at ($(0,-2*\dyb)$)         {$a_{M-1,n}$};
	\node[circ1,anchor=west]  (mul2) at ($(in2.east)  +(+\dx,+0.0)$)  {$\times$};
	\node[rect0,anchor=north] (phi2) at ($(mul2.south)+(+0.0,-\dya)$)  {$e^{j\phi_{M-1,n}}$};
	\node[rect1,anchor=west]  (exp2) at ($(mul2.east) +(+\dx,+0.0)$)  {$\uparrow\frac{M}{2}$};
	\node[rect2,anchor=west]  (f2)   at ($(exp2.east) +(+\dx,+0.0)$)  {$p_{M-1}[k]$};	
	\node[circ1,anchor=west]  (sum)  at ($(f1.east)   +(+\dx,+0.0)$)  {$+$};
	\node[rect1,anchor=west]  (db)   at ($(sum.east)  +(+1.6*\dx,+0.0)$)  {$z^{-\Delta_\beta}$};
    \draw[->] (in0)  -> node{} (mul0);
    \draw[->] (in1)  -> node{} (mul1);
    \draw[->] (in2)  -> node{} (mul2);
    \draw[->] (phi0) -> node{} (mul0);
    \draw[->] (phi1) -> node{} (mul1);
    \draw[->] (phi2) -> node{} (mul2);
    \draw[->] (mul0) -> node{} (exp0);
    \draw[->] (mul1) -> node{} (exp1);
    \draw[->] (mul2) -> node{} (exp2);
    \draw[->] (exp0) -> node{} (f0);
    \draw[->] (exp1) -> node{} (f1);
    \draw[->] (exp2) -> node{} (f2);
    \draw[->] (f0) -| node{} (sum);
    \draw[->] (f1) -> node{} (sum);
    \draw[->] (f2) -| node{} (sum);
    \draw[->] (sum) -> node[text width=0mm, pos=0.1, above left]{$s[k]$} (db);
	\node[dot1,anchor=west]   (nd)    at ($(db.east)  +(+\dx/2,+0.0)$)  {};
	\node[rect2,anchor=west]  (g0)    at ($(nd.east)   +(+\dx,+\dyb)$)   {$p_0[k]$};
	\node[rect1,anchor=west]  (d0)    at ($(g0.east)   +(+\dx,+0.0)$)   {$\downarrow\frac{M}{2}$};
	\node[circ1,anchor=west]  (mul0r) at ($(d0.east)   +(+\dx,+0.0)$)   {$\times$};
	\node[rect0,anchor=north] (phi0c) at ($(mul0r.south)+(+0.0,-\dya)$)  {$e^{-j\phi_{0,n-\Delta_\alpha}}$};
	\node[rect0,anchor=west]  (out0)  at ($(mul0r.east) +(+\dx,+0.0)$)   {$\tilde{a}_{0,n-\Delta_\alpha}$};
	\node[rect2,anchor=west]  (g1)    at ($(nd.east)   +(+\dx,+0.0)$)   {$p_1[k]$};
	\node[rect1,anchor=west]  (d1)    at ($(g1.east)   +(+\dx,+0.0)$)   {$\downarrow\frac{M}{2}$};
	\node[circ1,anchor=west]  (mul1r) at ($(d1.east)   +(+\dx,+0.0)$)   {$\times$};
	\node[rect0,anchor=north] (phi1c) at ($(mul1r.south)+(+0.0,-\dya)$)  {$e^{-j\phi_{1,n-\Delta_\alpha}}$};
	\node[rect0,anchor=west]  (out1)  at ($(mul1r.east) +(+\dx,+0.0)$)   {$\tilde{a}_{1,n-\Delta_\alpha}$};
	\node[rect2,anchor=west]  (g2)    at ($(nd.east)   +(+\dx,-\dyb)$)   {$p_{M-1}[k]$};
	\node[rect1,anchor=west]  (d2)    at ($(g2.east)   +(+\dx,+0.0)$)   {$\downarrow\frac{M}{2}$};
	\node[circ1,anchor=west]  (mul2r) at ($(d2.east)   +(+\dx,+0.0)$)   {$\times$};
	\node[rect0,anchor=north] (phi2c) at ($(mul2r.south)+(+0.0,-\dya)$)  {$e^{-j\phi_{M-1,n-\Delta_\alpha}}$};
	\node[rect0,anchor=west]  (out2)  at ($(mul2r.east) +(+\dx,+0.0)$)   {$\tilde{a}_{M-1,n-\Delta_\alpha}$};
	\draw[->] (db)  -> node{} (g1);
	\draw[->] (nd)  |- node{} (g0);
	\draw[->] (nd)  |- node{} (g2);
    \draw[->] (g0)  -> node{} (d0);
    \draw[->] (g1)  -> node{} (d1);
    \draw[->] (g2)  -> node{} (d2);
    \draw[->] (d0)  -> node{} (mul0r);
    \draw[->] (d1)  -> node{} (mul1r);
    \draw[->] (d2)  -> node{} (mul2r);
    \draw[->] (g0)  -> node{} (d0);
    \draw[->] (g1)  -> node{} (d1);
    \draw[->] (g2)  -> node{} (d2);
    \draw[->] (phi0c)  -> node{} (mul0r);
    \draw[->] (phi1c)  -> node{} (mul1r);
    \draw[->] (phi2c)  -> node{} (mul2r);
    \draw[->] (mul0r)  -> node{} (out0);
    \draw[->] (mul1r)  -> node{} (out1);
    \draw[->] (mul2r)  -> node{} (out2);
    \coordinate (nn1) at ($(db.west|-f2)+(-0.3,-1.1)$);
    \coordinate (nn2) at ($(db.west|-f0)+(-0.3,+3*\dya)$);
    \draw[dashed] (nn1) -- (nn2);
    \node[]  at ($(nn1)+(-1.5,0.1)$) {$\leftarrow$ Transmitter};
    \node[]  at ($(nn1)+(+1.5,0.1)$) {Receiver $\rightarrow$};
	\draw[decoration={brace,raise=5pt},decorate]($(mul0.east)+(0.25,1.2*\dya)$) -- node[above=5pt] {Synthesis Filter Bank} ++(4.5,0);
	\draw[decoration={brace,raise=5pt},decorate]($(g0.west)+(-0.7,1.2*\dya)$) -- node[above=5pt] {Analysis Filter Bank} ($(d0.east)+(0.25,1.2*\dya)$);
\end{tikzpicture}
\end{center}
\caption{\ac{FBMC} Transmultiplexer}
\label{fig:fbmc_diag1}
\end{figure*}

Fig. \ref{fig:fbmc_diag1} presents a complete FBMC transmultiplexer schematic. In this representation, PAM symbols from different subchannels are phase shifted, expanded at a rate of $M/2$, pulse-shaped by their respective subcarrier filter
\begin{equation}
	p_m[k] = p[k] e^{ j\frac{2\pi}{M} m \rbrac{k-\frac{L_p-1}{2} }}
\end{equation}
and combined to form the multiplexed signal $s[k]$. At the receiver side, the delay $\Delta_\beta$ ensures the causality of the scheme, while the latency $\Delta_\alpha$ arises at the estimated symbols. The relation between $\Delta_\alpha$ and $\Delta_\beta$ is given by
\begin{equation}
	L_p-1 = \dfrac{M}{2}\Delta_\alpha - \Delta_\beta,
\end{equation}
as demonstrated in \cite[sec. II.C]{Siohan2002}. Since $\Delta_\alpha$ and $\Delta_\beta$ are non-negative integers, it is easy to verify that a longer prototype filter introduces a higher latency at the output of the transmultiplexer. Thus, it is desirable to deploy filters with a low overlapping factor in order reduce the latency. However, desirable spectrum and reconstructions features usually come at the expense of longer prototype filters.

\section{Figures of Merit for Prototype Filters}\label{sec:fbmc_merit}
In this section, we present some figures of merit to evaluate the performance of prototype filters from different perspectives, which can be deployed to design or select the most suitable filter for a given application. The figures of merit herein presented {suitably characterize} prototype filters according to their spectrum leakage, time-frequency localization, and reconstruction capabilities, as {shown} in the {sequel}. 

\subsection{Signal-to-Interference Ratio}
According to eq. \eqref{eq:fbmc_est1}, the prototype filter impacts the symbol estimation at the receiver side. Hence, considering the scenario depicted in Fig. \ref{fig:fbmc_diag1}, the estimated symbols experience an \ac{SIR} expressed by
\begin{equation}
	\text{SIR} = \dfrac{1}
	                   { \displaystyle\sum_{m=0}^{M-1}
	                     \displaystyle\sum_{n=-\infty}^{\infty}
	                     \real{\left< p_{m,n}[k] \left\vert p_{m_0,n_0}[k] \right.\right>}^2
	                   },
	                   \label{eq:sir1}
\end{equation}
which can be used to quantize the symbol reconstruction quality. Typically, prototype filters are designed to provide an \ac{SIR} level of dozens of dBs, \textit{e.g.}, \cite[tab. I]{Mirabbasi2003}. Indeed, there are some other alternatives such as the maximum distortion parameter \cite[eq. (44)]{Siohan2002} and the prototype filter noise floor \cite[eq. (4)]{Bellanger2001}, which describe the self-interference of the prototype filter similarly to eq. \eqref{eq:sir1}.

\subsection{In and Out-of-Band Energy}
Another concern that arises when designing prototype filters for multicarrier applications is the amount of energy emitted outside the passband. Low \ac{OoB} energy emission ensures high energy efficiency and low interference to adjacent bands, which are desirable features a Cognitive Radio must comply when adapting itself for an opportunistic spectrum usage \cite{Kumar2016}.

The energy contained within the frequency range $\abs{\omega}\leq\omega_c$ can be defined as
\begin{equation}
	E(\omega_c) = \dfrac{1}{2\pi}
		           \displaystyle\int_{-\omega_c}^{\omega_c} 
		           \abs{P(e^{j\omega})}^2 d\omega,
	\label{eq:energy_ib1}
\end{equation}
where $P(e^{j\omega})$ is the \ac{DTFT} of $p[k]$. Typically, $\omega_c$ is set to $1/M$ as it is the subcarrier frequency separation.

More conveniently, eq. \eqref{eq:energy_ib1} may be expressed in matrix form, by defining the vector
\begin{equation}
	\mb{p}=
	\begin{bmatrix}
		p[0]     \quad 
		p[1]     \quad
		\cdots   \quad
		p[L_p-1]
	\end{bmatrix}^T
\end{equation}
and the entries of the matrix $\mb{\Gamma}(\omega_c)$ as \cite[eq. (3.2.18)]{Vaydianathan1993}
\begin{equation}
	\sbrac{\mb{\Gamma}(\omega_c)}_{k,l} = \dfrac{\omega_c}{\pi}\sinc\sbrac{(k-l)\dfrac{\omega_c}{\pi}}.
\end{equation}
Thus, the \ac{IB} energy can be evaluated through
\begin{equation}
	E(\omega_c) = \mb{p}^T \mb{\Gamma}(\omega_c) \mb{p},
	\label{eq:energy_ib2}
\end{equation}
while, the total energy of the filter can be expressed as
\begin{equation}
	E(\pi) = \mb{p}^T \mb{p}
\end{equation}
and the energy outside the frequency range $\abs{\omega}\leq\omega_c$, or \ac{OoB} energy, is evaluated by
\begin{equation}
	\overline{E}(\omega_c) = \mb{p}^T \sbrac{ \eye - \mb{\Gamma}(\omega_c) } \mb{p}.
	\label{eq:energy_ob2}
\end{equation}

\subsection{Maximum Sidelobe Level}
As suggested by the name, the \ac{MSL} measures the ratio between the maximum sidelobe of $\abs{P\rbrac{e^{j\omega}}}^2$ and the main lobe level. The MSL can be defined as
\begin{equation}
	\text{MSL} = \dfrac
{\displaystyle
	             \max_{\omega\in\mathbb{W}}
	             \abs{P(e^{j\omega})}^2}
	{\abs{P(e^{j0})}^2},
\end{equation}
where
\begin{equation}
\mathbb{W}
=
\cbrac{
\omega\in\mathbb{R } 
\left\vert 
\begin{array}{l}
\omega > 0, \quad
\dfrac{d}{d\omega}\abs{P(e^{j\omega})}^2=0 
\end{array}
\right. }.
\end{equation}
Notice that the \ac{MSL} describes the interference generated by $p[k]$ to adjacent bands.

\subsection{Heisenberg Factor}
As a {significant} amount of telecommunication systems operates under large 
{delay} and/or Doppler spreads, their pulse shaping is expected to be well localized in order to deal with such a harsh environment. Hence, it is highly desirable to deploy a prototype filter with low time and frequency 
{spreading}, which are defined respectively as
\begin{equation}
	D_k^2 = \displaystyle\sum_{k=-\infty}^{\infty}
	             \underline{k}^2 \abs{p[k]} ^2
\end{equation}
and
\begin{equation}
	D_\nu^2 = \displaystyle\int_{-1/2}^{1/2}
	          \nu^2  
	          \abs{P(e^{j2\pi\nu})}^2
	          d\nu.
\end{equation}

Unfortunately, a prototype filter cannot be designed to achieve an arbitrary time-frequency localization, as time  and frequency 
{spreadings} are conflicting goals. Indeed, this statement is known as the Heisenberg Uncertainty Principle, described by the inequality
\begin{equation}
	0 \leq \xi\leq 1,
\end{equation}
where
\begin{equation}
	\xi = \dfrac{1}{4\pi D_k D_\nu},
\end{equation}
which is referred as the Heisenberg parameter. Notice that well located pulses can achieve close to unit $\xi$, while poorly located pulses may achieve near null values of $\xi$.

\section{Prototype Filters}\label{sec:proto}
This section provides a brief background on some prototype filter options for \ac{FBMC} systems. In particular, the most popular prototype filter choices are the \ac{EGF}, Martin-Mirabbasi and the \ac{OFDP}, due to their remarkable features and simplicity. Nevertheless, other less popular options include the windowed based prototype filter \cite[eq. (34)]{Phydias}, the Hermite filter introduced in \cite[eq. (16)]{Haas1997} and the classical \ac{SRRC} \cite[eq. (24)]{Farhang2011}. Moreover, there are still some more recent developments in this field, which include the works \cite{Prakash2013} and \cite{Aminjavaheri2017}. 

As this paper does not aim to provide a full survey on prototype filters, this section offers a brief background on the \ac{EGF}, Martin-Mirabbasi and the \ac{OFDP} prototype filters, which are the best candidates for \ac{FBMC} systems to come \cite[sec. 7.2.1.1]{Afif2016}. In fact, there are already some works that provide a rich discussion on prototype filters, \textit{e.g.,} \cite{Sahin2014}.

\subsection{Extended Gaussian Function}
Originally, the \ac{EGF} was generated using the \ac{IOTA} \cite[eq. (25)]{Lefloch1995}
\begin{equation}
	\mathcal{O}_a x(t) = \dfrac{x(t)}{\sum_{-\infty}^{\infty} \abs{ x(t-a i) }^2}
\end{equation}
on a Gaussian function with a spreading factor $\alpha$, \textit{i.e.,} $g_\alpha(t)=(2\alpha)^{1/4}  e^{ -\pi \alpha t^2 }$. Hence, the continuous \ac{EGF} is defined as
\begin{equation}
	p(t) = \mathcal{F}^{-1}\mathcal{O}_{\tau_0} \mathcal{F}  \mathcal{O}_{\nu_0} g_{\alpha}(t),
	\label{eq:fbmc_egf_cont}
\end{equation}
where $\mathcal{F}$ is the Fourier transform operator, $\tau_0$ is the signaling interval and $\nu_0$ is the subcarrier spacing, which must comply $\tau_0\nu_0=1/2$ for \ac{FBMC} systems. In order to obtain the discrete version of an \ac{EGF}, one can sample it properly. Fortunately, an analytic expression for \eqref{eq:fbmc_egf_cont} is provided in \cite[eq. (7)]{Siohan2000}.

As the \ac{IOTA} aims to make \ac{EGF} pulses orthogonal, they are expected to provide a high quality symbol reconstruction. In fact, \ac{SIR} level provided by an \ac{EGF} can be adjusted by tuning the spread factor $\alpha$, where the \ac{SIR} is proportional to $\alpha$, as can be observed in \cite[fig. 3]{Siohan2000}. However, by increasing $\alpha$, the \ac{EGF} pulse becomes shorter in time and, thus, a higher frequency dispersion is expected. 

\subsection{Mirabbasi-Martin}
To ensure fast spectrum decay throughout the stopband region, the Mirabbasi-Martin prototype filter \cite{Mirabbasi2003} focuses on minimizing the discontinuity in their boundaries, while maintaining good reconstruction features for multicarrier applications and ensuring a smooth pulse variation. This design uses the frequency sampling technique, where the filter weights are actually samples of the frequency response of the prototype filter.

Due to its fast spectrum decay and good performance for data reconstruction, Mirabbasi-Martin prototype is the main choice for the Phydias project \cite{Phydias}, which aims to enable \ac{FBMC} applications in wireless systems to come. Indeed, some authors refer such a filter, for $K=4$, as the PHYDIAS filter. 

The Mirabbasi-Martin prototype filter can be written as the following discrete low-pass filter:
\begin{equation}
\resizebox{.43\textwidth}{!}
{
\text
{
$
\arraycolsep=1.4pt
p[k] =
\left\lbrace
\begin{array}{ll}
	k_0 + 2\displaystyle\sum_{i=1}^{K-1}  k_i \cos\rbrac{ \dfrac{2\pi i}{KM}k }, & 0\leq k \leq L_p-1\\
	0, & \text{otherwise}
\end{array}
\right.
\label{eq:martin_discrete}
$
}
}
\end{equation}
where $k_\ell$ are the filter weights which are available in \cite[tab. I]{Mirabbasi2003}.

\subsection{Prolate Filter and Discrete Slepian Sequences}\label{sec:prolate}
The Prolate filter is a classic design that aims to maximize the energy within its passband region. This design can be compactly expressed as
\begin{equation}
	\pmb{\psi}_{0,\omega_s} = 
	\begin{array}[t]{cl}
		\text{argmax} & \mb{p}^H \mb{\Gamma}(\omega_s) \mb{p}\\[3mm]
		\text{s.t.}      & \mb{p}^H\mb{p}
		                  = 1
	\end{array}.
	\label{eq:prolate_p2}
\end{equation}
Since $\mb{\Gamma}(\omega_s)$ is symmetric, a straightforward solution comes by recalling the Rayleigh-Ritz Theorem \cite[theo. 4.4.2]{Horn1985}, which guarantees the solution of \eqref{eq:prolate_p2} to be the eigenvector associated to the largest eigenvalue of $\mb{\Gamma}(\omega_s)$. By denoting $\gamma_0\geq\gamma_1\geq\cdots\geq\gamma_{L_p-1}>0$ the eigenvalues of $\mb{\Gamma}(\omega_s)$, and $\pmb{{\psi}}_{i,\omega_s}$ the eigenvector associated to $\gamma_i$, the solution of \eqref{eq:prolate_p2} is
\begin{equation}
	\pmb{\psi}_{0,\omega_s}=
	\begin{bmatrix}
		\psi_{0,\omega_s}[0]     \quad
		\psi_{0,\omega_s}[1]     \quad
		\cdots   	  \quad
		\psi_{0,\omega_s}[L_p-1]
	\end{bmatrix}^T.
\end{equation}

Physically, $\gamma_i$ represents the normalized energy of $\psi_i[k]$ within $\abs{\omega}\leq\omega_c$, thus
\begin{equation}
	0\leq\gamma_i\leq1.
\end{equation}
Consequently, the sequence $\psi_{0,\omega_s}[k]$ is the most selective filter for {the problem stated in eq.} \eqref{eq:prolate_p2}. It is also noteworthy mentioning that, the remaining filters $\psi_{i,\omega_s}[k]$ are local solutions, possessing less energy within their passband than $\psi_{0,\omega_s}[k]$. Indeed, $\cbrac{\psi_{i,\omega_s}[k]}_i$ is also known as the \ac{DPSS} or the Slepian series \cite{Slepian1978}. Interestingly, the Slepian series is a good choice to interpolate smooth functions \cite[tab. 2]{Moore2004}, making it suitable for low \ac{OoB} emission {applications}.

\subsection{Optimal Finite Duration Pulse}
As stated previously, the Prolate design is optimal in terms of minimizing the energy outside the passband. However, (near) perfect reconstruction requirements are not taken into account in this design. From this perspective, the \ac{OFDP} deploys the Slepian series to provide a filter design with low \ac{OoB} emission and a good symbol reconstruction capability. The \ac{OFDP} can be written as
\begin{equation}
	p[k] = \displaystyle\sum_{i} \alpha_{2i} \psi_{2i,2\pi/M}[k], 
	\label{eq:ofdp_seq}
\end{equation}
where the coefficients $\alpha_{2i}$ can be found in \cite[tab. I]{Vahlin1994}. Since the \ac{IB} energy of $\psi_{2i}$, \textit{i.e.}, $\gamma_{2i}$, decays rapidly as can be observed in \cite[fig. 3,4]{Slepian1978}, truncation is acceptable for solving the problem.

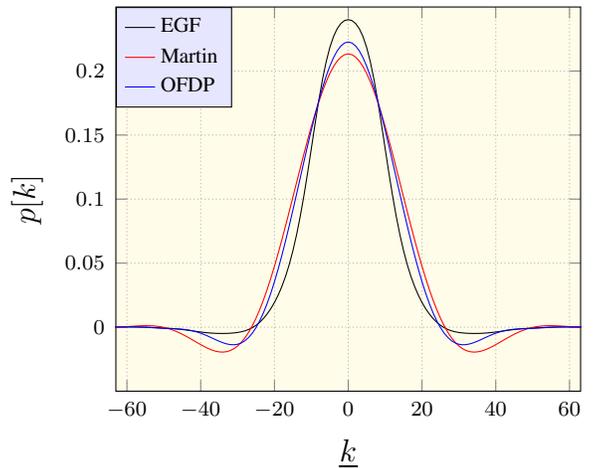
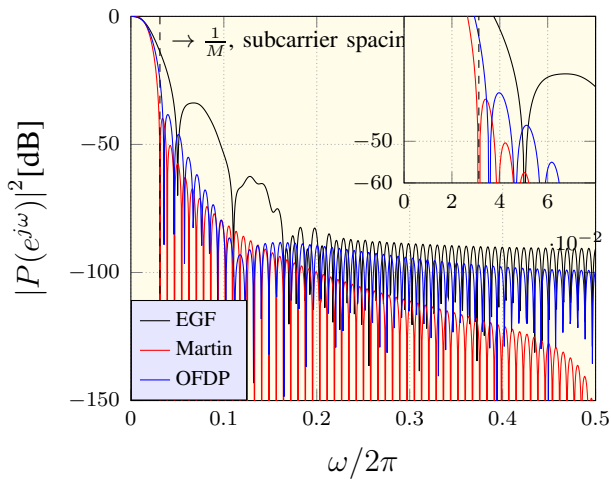
\begin{figure}[!htbp]
	\centering
	\subfloat[Impulse Response\label{fbmc_pulse_time_pop_K=4}]
	{\begin{tikzpicture}[scale=1]
		\begin{axis}[ axisstyle,
					  legend columns=1,legend style={at={(0,1)},anchor=north west},
		              ylabel={$p[k]$}, xlabel={$\underline{k}$}, 
		              ytick={0,.05,.1,.15,.20},
		              xmin=-63,xmax=+63, ymin=-.05,ymax=.25,
		              mark repeat=4]
		    \legend{EGF,Martin,OFDP}
			\addplot[egf,    mark=none] table[x index=0,y index=1] {fbmc_pulse_time_egf_K=4_M=32.dat};
			\addplot[martin, mark=none] table[x index=0,y index=1] {fbmc_pulse_time_martin_K=4_M=32.dat};
			\addplot[ofdp,   mark=none] table[x index=0,y index=1] {fbmc_pulse_time_ofdp_K=4_M=32.dat};
		\end{axis}
	\end{tikzpicture}}
	\\
	\subfloat[Frequency Response\label{fbmc_pulse_freq_pop_K=4}]
	{\begin{tikzpicture}[scale=1]
		\begin{axis}[ axisstyle,
		              legend columns=1,
		              legend style={at={(0,0)},anchor=south west},
		              ylabel={$\abs{P(e^{j\omega})}^2$[dB]}, xlabel={$\omega/2\pi$}, 
		              xmin=+0.0,xmax=+0.5,
		              ymin=-150,ymax=0]
		    \legend{EGF,Martin,OFDP}
			\addplot[egf,    mark=none] table[x index=0,y index=1] {fbmc_pulse_freq_egf_K=4_M=32.dat};
   			\addplot[martin, mark=none] table[x index=0,y index=1] {fbmc_pulse_freq_martin_K=4_M=32.dat};
			\addplot[ofdp,   mark=none] table[x index=0,y index=1] {fbmc_pulse_freq_ofdp_K=4_M=32.dat};		
			\addplot[color=black,dashed] coordinates{(1/32,0) (1/32,-300)};
			\node[anchor=north west] at (axis cs: 0.03125,0) {\footnotesize $\rightarrow\frac{1}{M}$, subcarrier spacing};
			\coordinate (posx) at (rel axis cs:1,1);
		\end{axis}
		\begin{axis}[ axisstyle,at={(posx)},anchor={north east},
		              width  = .25\textwidth, 
		              height  = .23\textwidth,
    	              xtick={0,.02,.04,.06},
    				  ytick={-50,-60,-70},
    	              xmin=+0.00,xmax=+.08,
    	              ymin=-60,ymax=-20,
    	              ]
			\addplot[egf,    mark=none] table[x index=0,y index=1] {fbmc_pulse_freq_egf_K=4_M=32.dat};
   			\addplot[martin, mark=none] table[x index=0,y index=1] {fbmc_pulse_freq_martin_K=4_M=32.dat};
			\addplot[ofdp,   mark=none] table[x index=0,y index=1] {fbmc_pulse_freq_ofdp_K=4_M=32.dat};	
			\addplot[color=black,dashed] coordinates{(1/32,0) (1/32,-300)};
    	\end{axis}
	\end{tikzpicture}}
	\caption{Prototype filters comparison for $K=4$ and $M=32$.}
	\label{fig:fbmc_pulse_pop_K=4_M=32}
\end{figure}

As a preview, Fig. \ref{fig:fbmc_pulse_pop_K=4_M=32} depicts the impulse and frequency responses of the \ac{EGF}, Martin and \ac{OFDP} prototype filters. Qualitatively, one can observe that Martin prototype {filter} presents the fastest spectrum decay, reaching approximately $-150$[dB] at $\omega=\pi$, while both \ac{OFDP} and \ac{EGF} spectrum floor is around $-100$[dB]. Also, the \ac{EGF} with $\alpha=1$ is shorter in time, leading to a higher frequency dispersion. A more detailed discussion is provided throughout the numerical results, along with a comparison with the proposed filter design.

\section{Proposed Design}\label{sec:fbmc_proposed_design}
In this section, we propose a prototype filter design methodology based on convex optimization. Through this design, we aim to minimize the \ac{OoB} energy emission, while providing a high quality symbol reconstruction and maintaining a fast spectrum decay. The description of the proposed design begins by defining the filter expression as a linear transformation. In the sequence, we provide the objective function expression, \textit{i.e.}, the \ac{OoB} energy. Furthermore, we present a full discussion on the \ac{FBMC} interference elements, which is be used to ensure a high \ac{SIR} prototype filter. The constraints required to achieve a prototype filter with fast spectrum decay are also offered. Finally, we cast the complete problem as a non-convex \ac{QCQP}, which on its turn is relaxed into a convex \ac{QCQP} guided by a line search. Since our design depends on convex optimization, we present the design itself alongside with all the convexity proofs. This choice is  made aiming to favor the comprehension of both the deployed optimization method and the related convexity issues.

\subsection{Filter Expression}
In order to model the prototype filter, let us define the matrix
\begin{equation}
	\mb{F} = 
	\begin{bmatrix}
		\mb{f}_0 \quad
		\mb{f}_2 \quad
		\cdots \quad
		\mb{f}_{N-1}
	\end{bmatrix}.
\end{equation}
to be an aggregation of $N$ sequences, where
\begin{equation}
	\mb{f}_i=
	\begin{bmatrix}
		f_i[0]     \quad
		f_i[1]     \quad
		\cdots   	  \quad
		f_i[L_p-1]
	\end{bmatrix}^T
\end{equation}
is a vector with unitary norm $\norm{\mb{f}_i}_2=1$. Hence, let us express the prototype as the linear transformation
\begin{equation}
	\mb{p} = \mb{F} \mb{c},
	\label{eq:opt_lin_trans}
\end{equation}
where
\begin{equation}
	{\bf c} =
	\begin{bmatrix}
		c_0 \quad
		c_2 \quad
		\cdots    \quad
		c_{N-1}
	\end{bmatrix}^{T}
\end{equation}
are the coefficients to be optimized. 

Throughout this paper, we consider two families to be deployed as $f_i[k]$\footnote{One can deploy other sequences as long as they are (near) orthogonal, symmetrical around $(L_p-1)/2$.}. First, we consider $f_i[k]$ as the \ac{DPSS}, \textit{i.e.}, 
\begin{equation}
	f_i[k] = \psi_{2i,\omega_s}[k].
	\label{eq:fbmc_basis_slepian}
\end{equation}
As an alternative, a family of cosine sequences can also be deployed:
\begin{equation}
	\hspace*{-2mm}
	f_i[k] = 
	\left\lbrace
	\hspace*{-1.5mm}
	\begin{array}{ll}
		\dfrac{1}{\sqrt{KM+1}}, & \hspace*{-1.5mm} i=0 \\[10pt]
		\sqrt{\dfrac{2}{KM+2}} 
		\cos\rbrac{ \dfrac{2\pi i}{KM} \underline{k} }, & \hspace*{-1.5mm} i=1,\cdots N-1
	\end{array}
	\right. \hspace*{-3mm}.
	\label{eq:fbmc_basis_cosine}
\end{equation}
Notice that, large values of $N$ may increase the \ac{OoB} frequency content or make the last entries of $\mb{c}$ very small. Thus, since $N<L_p$, the number of optimization variables is considerably reduced.

\subsection{Energy Expression}
Combining eq. \eqref{eq:energy_ob2} and \eqref{eq:opt_lin_trans}, one may rewrite the energy concentrated within the stopband region as
\begin{equation}
	\arraycolsep = 1.4pt
	\begin{array}{rcl}
		\overline{E}(\omega_c) & = & \mb{c}^T \sbrac{ \eye_{L_p} - \mb{\Gamma}(\omega_c) } \mb{c} \\[3pt]
		                       & = & \mb{c}^T \mb{Q}_0                                     \mb{c}
	\end{array}
	\label{eq:objective_function}
\end{equation} 
which is an appropriate choice as it is a quadratic convex function, given $\mb{Q}_0$ is a positive semidefinite matrix. A more straightforward, yet informal, way of proofing that $\mb{Q}_0$ is positive semidefinite is {by recalling} that $\overline{E}(\omega_c)$ is an energy measurement, \textit{i.e.}, a non-negative value. Thus, if $\overline{E}(\omega_c)$ is non-negative, ${\mb{Q}_0}$ is positive semidefinite. Since ${\mb{Q}_0}$ is positive semidefinite, eq. \eqref{eq:objective_function} is convex \cite[sec. 1]{Lu2011}.

\subsection{\ac{FBMC} Interference Elements}
As observed in eq. \eqref{eq:fbmc_est2}, the quality of the symbol reconstruction depends on the prototype filter, which needs to be designed to provide low distortion levels and enable near-perfect reconstruction. To simplify the representation of the interference for our problem, let us define
\begin{equation}
	\arraycolsep = 1.4pt
	\hspace*{-2.5mm}
	\begin{array}{rcl}
		\epsilon_{m,n} & = & \real{\left< p_{m,n}[k] \left\vert p_{0,0}[k] \right.\right>}\\[3pt]
					   & = & \cos\rbrac{ \phi_{m,n} } 
					         \hspace*{-1mm}
		                   	 \displaystyle\sum_{k=-\infty}^{\infty}
		                   	 \hspace*{-1mm}
							 p\sbrac{k-n\frac{M}{2}} p[k]
					         \cos\rbrac{ \dfrac{2\pi}{M} m  \underline{k}  }
	\end{array}
	\hspace*{-3mm}
	\label{eq:fbmc_dist}
\end{equation}
as the distortion or interference introduced by the symbol $a_{m,n}$ {into} the symbol $a_{0,0}$. 

Based on \eqref{eq:fbmc_dist}, five straightforward properties can be listed:
\begin{itemize}
	\item[i]   $\epsilon_{0,0}$ represents the pulse energy;
	\item[ii]  $\epsilon_{m,n}$ is an even sequence concerning the index $n$, \textit{i.e.}, $\epsilon_{m,n}=\epsilon_{m,-n}$;
	\item[iii] $\epsilon_{m,n}$ is odd circular symmetric concerning the index $m$, \textit{i.e.}, $\epsilon_{m,n}=-\epsilon_{M-m,n}$, for $1\leq m\leq M/2-1$;
	\item[iv]  $\epsilon_{m,n}=0$ for $\abs{n}>\ceil{\frac{L_p}{M/2}}-1$, since the prototype filter is finite;
	\item[v]   $\epsilon_{m,n}=0$ case $m+n$ is odd, given the cosine term.
\end{itemize}
Thus, taking into account properties ii, iii and iv, let us define 
\begin{equation}
	\mathcal{E} = 
	\cbrac
	{
	(m,n)
	\left\vert
	\begin{array}{l}
		0\leq m \leq M/2 \\
		0\leq n \leq \ceil{\frac{Lp-1}{M/2}}-1 \\
		m+n \text{ even} \\
		m+n\neq 0
	\end{array}
	\right.
	}
\end{equation}
to be the set of distortion elements $\epsilon_{m,n}$ which are not strictly null, but can assume a negligible values depending on the prototype filter design. Given the symmetry of $\epsilon_{m,n}$, we omitted the redundant elements in order to propose an efficient optimization problem.

\subsubsection{Matrix Form}
In order to model the interference elements $\epsilon_{m,n}$ in matrix form, two matrices need to be defined. First, 
 \begin{equation}
	[\mb{\Pi}_{n}]_{i,j} = 
	\begin{cases}
		1, \quad j=i+nM/2\\
		0, \quad \text{otherwise}
	\end{cases}
	\label{eq:shift_matrix}
\end{equation}
are the entries of the nilpotent matrix $\mb{\Pi}_{n}$, responsible for shifting the sequence $p[k]$ by $nM/2$ samples. The second matrix, $\mb{\Sigma}_m$, incorporates the cosine term of the summation of eq. \eqref{eq:fbmc_dist} and is defined as
\begin{equation}
	\mb{\Sigma}_m = \text{diag}\cbrac{ \cos\rbrac{\dfrac{2\pi}{M}m\underline{k}} }_{k=0,1,\cdots, L_p-1}.
	\label{eq:cosine_matrix}
\end{equation}

Initially, we can rewrite eq. \eqref{eq:fbmc_dist} in matrix form:
\begin{equation}
	\arraycolsep=1.4pt
	\begin{array}{rcl}
	\epsilon_{m,n} & = & \mb{p}^T { \cos(\phi_{m,n})\mb{\Sigma}_m \mb{\Pi}_{n} }\mb{p}\\[3pt]
	               & = & \mb{p}^T \mb{Q}_{m,n}^{(0)} \mb{p}.
	\end{array}
	\label{eq:fbmc_interf1}
\end{equation}
Notice that, since $\mb{\Sigma}_m$ is diagonal and $\mb{\Pi}_n$ is nilpotent, $\mb{Q}_{m,n}^{(0)}$ is also nilpotent for $n\neq0$. Conversely, case $n=0$, the eigenvalues of $\mb{Q}_{m,n}^{(0)}$ lies on the diagonal of $\mb{\Sigma}_m$, as $\mb{\Sigma}_m$ is diagonal and $\mb{\Pi}_0=\eye_{L_p}$. Thus, $\mb{Q}_{m,n}^{(0)}$ is not a set of positive semidefinite matrices making $\epsilon_{m,n}$ to be a non-convex set.

For convenience, $\epsilon_{m,n}$ can be expressed using Hermitian symmetric matrices. Thus, by noting that $\epsilon_{m,-n}=\mb{p}^T \rbrac{\mb{Q}_{m,n}^{(0)}}^{T} \mb{p}$ and $\epsilon_{m,-n}=\epsilon_{m,n}$, eq. \eqref{eq:fbmc_dist} can be expressed as
\begin{equation}
	\arraycolsep=1.4pt
	\begin{array}{rcl}
	\epsilon_{m,n} & = & \dfrac{1}{2} \mb{p}^T \sbrac{\mb{Q}_{m,n}^{(0)} + \rbrac{\mb{Q}_{m,n}^{(0)}}^T}\mb{p}\\[8pt]
	               & = &              \mb{p}^T \mb{Q}_{m,n}^{(1)} \mb{p}.
	\end{array}
\end{equation}
Since $\mb{Q}_{m,n}^{(1)}$ is a Hermitian matrix, it yields real eigenvalues. Unfortunately, to our knowledge, the exact eigenvalues of $\mb{Q}_{m,n}^{(1)}$, given $n\neq0$, cannot be tracked analytically. However, one can observe  that
\begin{equation}
	\abs{ \sum_\ell \sbrac{ \mb{Q}_{m,n}^{(1)} }_{i,\ell} }\leq1, \quad i\neq\ell.
\end{equation}
Hence, the Gershgorin Theorem \cite[Theorem 6.1.1]{Horn1985} guarantees that the eigenvalues are inside the interval
\begin{equation}
\rho\rbrac{\mb{Q}_{m,n}^{(1)}}\in[-1,1]
\end{equation}

In terms of the coefficients $c_i$, the interference element $\epsilon_{m,n}$ can be written as
\begin{equation}
	\arraycolsep=1.4pt
	\begin{array}{rcl}
	\epsilon_{m,n} & = & \mb{c}^T \mb{F}^T \mb{Q}_{m,n}^{(1)} \mb{F}\mb{c}\\[3pt]
	               & = & \mb{c}^T          \mb{Q}_{m,n}^{(2)} \mb{c}.
	\end{array}
\end{equation}
Again, $\mb{Q}_{m,n}^{(2)}$ is Hermitian and with eigenvalues that are not analytically {traceable}. Nevertheless, one can also realize that $\rho\rbrac{\mb{Q}_{m,n}^{(2)}}\in[-1,1]$, by analyzing the numerical radius of $\mb{Q}_{m,n}^{(2)}$.

\subsubsection{Self-Interference Constraints}
In order to manage the self-interference level generated by the prototype filter, the proposed problem is constrained to a maximum interference level of $\epsilon_0$
\begin{equation}
	\abs{\epsilon_{m,n}} =  \abs{ \mb{c}^T\mb{Q}_{m,n}^{(2)}\mb{c} }  \leq \epsilon_0, \quad  (m,n)\in \mathcal{E}.
	\label{eq:const_a}
\end{equation}
{Notice that} the absolute value {must be taken, since}  $\epsilon_{m,n}$ can assume both real and negative values. Alas, the constraint presented in \eqref{eq:const_a} is non-convex since $+\mb{Q}_{m,n}^{(2)}$ and $-\mb{Q}_{m,n}^{(2)}$ cannot be positive semidefinite simultaneously.

\subsubsection{Eigenvalues Shift}
The eigenvalues of $\mb{Q}_{m,n}^{(2)}$ can be manipulated  by adding the term $\delta\mb{c}^T\mb{F}^T\mb{F}\mb{c}$ and subtracting $\delta$ from eq. \eqref{eq:const_a}, leading to
\begin{equation}
	\begin{cases}
		{ \mb{c}^T\sbrac{ +\mb{Q}_{m,n}^{(2)} + \delta \mb{F}^T\mb{F} } \mb{c} } - \delta \leq \epsilon_0\\
		{ \mb{c}^T\sbrac{ -\mb{Q}_{m,n}^{(2)} + \delta \mb{F}^T\mb{F} } \mb{c} } - \delta \leq \epsilon_0
	\end{cases}\hspace*{-3mm},
	\hspace*{2mm}  
	(m,n)\in \mathcal{E},
	\label{eq:const_c}
\end{equation}
which is valid if and only if the energy of the prototype filter is unitary, \textit{i.e.}, $\norm{\mb{p}}_2^2=\mb{c}^T\mb{F}^T\mb{F}\mb{c}=1$. In order to provide a more compact notation, let us rewrite eq. \eqref{eq:const_c} as
\begin{equation}
	\begin{cases}
		{ \mb{c}^T \mb{Q}_{m,n}^{(3a)}  \mb{c} } \leq \epsilon_0 + \delta\\
		{ \mb{c}^T \mb{Q}_{m,n}^{(3b)}  \mb{c} }  \leq \epsilon_0 + \delta
	\end{cases},
	\label{eq:const_d}
\end{equation}

Considering the eigenvalues of $\mb{Q}_{m,n}^{(2)}$ and that both $\mb{F}^T \mb{F}$ and $\mb{Q}_{m,n}^{(2)}$ are Hermitian, one can easily observe that 
\begin{equation}
\hspace*{-2mm}
\rho\rbrac{\mb{Q}_{m,n}^{(3a)}} 
\in 
\sbrac{
\delta\rho_{\min}\rbrac{\mb{F}^T\mb{F}}-1,
\delta\rho_{\max}\rbrac{\mb{F}^T\mb{F}}+1
}
\label{eq:weyl1}
\end{equation}
by recalling the Weyl's Inequality. A particular case takes place if $\mb{F}$ is orthonormal, where
\begin{equation}
\rho\rbrac{\mb{Q}_{m,n}^{(3a)}} 
\in 
\sbrac{
\delta-1,
\delta+1
},
\label{eq:weyl2}
\end{equation}
which is the case if $\mb{F}$ is a Slepian basis, \textit{i.e.}, eq. \eqref{eq:fbmc_basis_slepian}. However, if $\mb{F}$ is composed by a cosine sequences, as the ones in eq. \eqref{eq:fbmc_basis_cosine}, such matrix is only near orthogonal, making the eigenvalues of $\mb{F}^T\mb{F}$ very close to the unit. Hence, there exists a non-negative value of $\delta$ which makes all the eigenvalues of $\mb{Q}_{m,n}^{(3a)}$ positive, making this matrix positive semidefinite. Similarly, we can also conclude that $\mb{Q}_{m,n}^{(3b)}$ is also positive definite for a given value of $\delta$. Hence, both entries of eq. \eqref{eq:const_d} are convex, as both $\mb{Q}_{m,n}^{(3a)}$ and $\mb{Q}_{m,n}^{(3b)}$ are positive semidefinite \cite{Boyd2004}.

\subsection{Spectrum Decay}
As practical \ac{FBMC} systems deploy windowed pulses, spectrum decay may stagnate, specially if the boundaries of the prototype filter are not null. Thus, the samples at the boundaries of the prototype filter should assume values as low as possible to ensure a fast spectrum decay. In this sense, let us define
\begin{equation}
	p[k] = \mb{c}^T \mb{u}_{k},
\end{equation}
where
\begin{equation}
	\mb{u}_{k} =
	\begin{bmatrix}
		f_0[k]     \quad
		f_1[k]     \quad
		\cdots   	  \quad
		f_{N-1}[k]
	\end{bmatrix}^T.
\end{equation}
Thus, the boundaries of the prototype filter are constrained by
\begin{equation}
	\abs{ \mb{c}^T \mb{u}_{k} } \leq u_0, \quad k\in\mathcal{K},
	\label{eq:constraint_boundary}
\end{equation}
where $\mathcal{K}$ is the set of indexes of the boundaries samples we wish to limit to a maximum level $u_0$. Since the prototype filter is symmetrical, \textit{i.e.}, $p[k]=p[Lp-1-k]$, one does not need to constraint the amplitude of the pulse from both sides.

\subsection{Resulting Optimization Problem}
By taking eq. \eqref{eq:objective_function} as the objective function and eqs. \eqref{eq:const_d} and \eqref{eq:constraint_boundary} {as constraints}, the resulting problem can be written:
\begin{equation}
	\begin{array}{rll}
		\mb{c}^* = 
		\text{argmin}   &  \mb{c}^T\mb{Q}_0\mb{c}
		                \\[3mm]
		 \text{s.t.}     & \mb{c}^T \mb{Q}_{m,n}^{(3a)}  \mb{c} \leq \epsilon_0 + \delta, 
		                & (m,n) \in \mathcal{E} 
		                \\[3mm]
		                & \mb{c}^T \mb{Q}_{m,n}^{(3b)}  \mb{c}  \leq \epsilon_0 + \delta, 
		                & (m,n) \in \mathcal{E} 
		                \\[3mm]
		                & \abs{\mb{u}_k^T \mb{c}} \leq u_0,
		                & k\in\mathcal{K}
		                \\[3mm]
		                & \mb{c}^T{\mb{F}^T\mb{F}} \mb{c}=1
	\end{array}
	\label{eq:prob_qcqp1}
\end{equation}
Unfortunately, the problem posed in \eqref{eq:prob_qcqp1} is non-convex as the energy equality constraint is quadratic instead of affine \cite{Boyd2004}.

\subsection{Convex \ac{QCQP} Relaxation}
As an alternative to circumvent the non-convexity {of} eq. \eqref{eq:prob_qcqp1}, let us propose a relaxation in order to obtain a convex \ac{QCQP}. First, consider the norm inequality
\begin{equation}
	\norm{\mb{c}}_2 \leq \norm{\mb{c}}_1 \leq \sqrt{N} \norm{\mb{c}}_2.
	\label{eq:norm_const}
\end{equation}
One can observe that there is a value of $\zeta$ such as
\begin{equation}
	\norm{\mb{c}}_2 = \dfrac{\norm{\mb{c}}_1}{\zeta}.
	\label{eq:norm_const2}
\end{equation}
By {analyzing} \eqref{eq:norm_const}, one can observe that eq. \eqref{eq:norm_const} is valid for
\begin{equation}
1\leq\zeta\leq \sqrt{N}
\end{equation}

Taking the previous observation, we propose 
{exchanging} the norm-2 equality constraint by a norm-1 equality 
\begin{equation}
\dfrac{\norm{\mb{c}}_1}{\zeta} = \dfrac{ \mb{1}^T \mb{c} }{\zeta}, 
\label{eq:norm_const3}
\end{equation}
where $\mb{1}$ is a vector of ones. Notice that, eq. \eqref{eq:norm_const3} holds if $c_i\geq0${,} { which can be obtained by a proper choice of $\mb{F}$. As a rule of thumb,  $f_i[k]$ is chosen to be symmetrical around $(L_p-1)/2$, i.e., 
\begin{equation}
	\sbrac{\mb{F}}_{\ell,i}=\sbrac{\mb{F}}_{L_p+1-\ell,i}, 
	\quad
	1 \leq \ell \leq \floor{\dfrac{L_p}{2}}, 
\end{equation}
where $f_i[k] = \sbrac{\mb{F}}_{k+1,i}$ for $0\leq k\leq L_{p}-1$. Also, the central sample of 
{${\bf f}_i$} must be  positive:
\begin{equation}
		\sbrac{\mb{F}}_{\ceil{\frac{L_p}{2}},i}>0
\end{equation}
Such features can be observed, for example, in Martin prototype filter, \ac{OFDP} and Hermite  designs if their respective functions are properly scaled. }

Therefore, a relaxed \ac{QCQP} can be cast from eq. \eqref{eq:prob_qcqp1} and \eqref{eq:norm_const3}:
\begin{equation}
	\hspace*{-2mm}
	\begin{array}{rll}
	{\mb{c}}(\zeta) = 
		\text{argmin}   &  \mb{c}^T\mb{Q}_0\mb{c}
		                \\[3mm]
		 \text{s.t.}     & \mb{c}^T \mb{Q}_{m,n}^{(3a)}  \mb{c} \leq \epsilon_0 + \delta , 
		                & (m,n) \in \mathcal{E} 
		                \\[3mm]
		                & \mb{c}^T \mb{Q}_{m,n}^{(3b)}  \mb{c}  \leq \epsilon_0 + \delta , 
		                & (m,n) \in \mathcal{E} 
		                \\[3mm]
		                & \abs{\mb{u}_k^T \mb{c}} \leq u_0,
		                & k\in\mathcal{K}
		                \\[3mm]
		                & \dfrac{\mb{1}^T\mb{c}}{\zeta} = 1
		                \\[3mm]
		                & \mb{c}\geq\mb{0}.
	\end{array}
	\label{eq:prob_qcqp2}
\end{equation}
The convexity of \eqref{eq:prob_qcqp2} can be guaranteed as the Hessian of both the objective function and the constraints are positive semidefinite, and, also, the equality constraint are affine \cite{Luo2010}. Nevertheless, one must first track the value of $\zeta^*$, for which $\norm{\mb{p}}_2$ is as close as possible to the unity, leading to a near optimal solution. In this sense, the {associated} line search {problem}
\begin{equation}
	\begin{array}{rl}
	\zeta^* = 
		\text{argmin}   &  \abs{1-\mb{c}^T(\zeta)\mb{F}^T\mb{F}\mb{c}(\zeta)}^2
		                \\[3mm]
		 \text{s.t.}     & 1\leq \zeta\leq \sqrt{N}
	\end{array}
	\label{eq:prob_qcqp2a}
\end{equation}
can be performed to track the optimal value of $\zeta$. One can observe that the line search posed in \eqref{eq:prob_qcqp2a} requires the solution/evaluation of eq. \eqref{eq:prob_qcqp2}. Also, the objective function of \eqref{eq:prob_qcqp2a} must reach very small values 
 {to} satisfy $\norm{\bf p}_2=1$. Hence, {once} $\zeta^*$ is found, the relaxed solution
\begin{equation}
	\tilde{\mb{c}}^* = \mb{c}(\zeta^*)
	\label{eq:prob_qcqp2b}
\end{equation}
is established.

\section{Numerical Results}\label{sec:numerical}
In this section, we offer the numerical results 
to corroborate the effectiveness of the proposed filter design methodology. First, we define the optimization setup for three proposed prototype filters. After that, we provide a brief explanation on the tools deployed to solve the proposed optimization problem. Finally, we offer a performance comparison between the filters obtained through the proposed methodology {\it versus} 
the \ac{EGF}, \ac{OFDP} and Martin {prototype filters}. 

\subsection{Optimization Setup}
In order to exemplify the effectiveness of the proposed design, we solve \eqref{eq:prob_qcqp2a} using three different configurations referred hereafter as Type-I, Type-II and Type-III {prototype filters}. All the three set of parameters are summarized in Table \ref{tab:fbmc_par}. It is noteworthy mentioning that we choose an overlapping factor $K=4$, as it enables achieving filters with a high performance for both \ac{SIR} and spectrum measurements. Furthermore, $M=32$ was chosen to enable a better spectrum visualization. However, we must highlight at this point that the proposed design is also capable of handling other values of $M$.
\begin{table}[!ht]
	\centering
	\caption{Optimization parameters setup for the proposed method considering $K=4$, $M=32$ and $L_p=KM+1$}
	\vspace*{-2mm}
	\scalebox{.9}
	{
	\setlength{\tabcolsep}{3pt}
	\renewcommand{\arraystretch}{1.5}
	\begin{tabular}{cccc}
		\hline
		\rowcolor{blue!10}
		\textbf{Parameter} & Type-I  & Type-II  & Type-III
		\\\hline
		
		\rowcolor{yellow!10}
		$N$               & $2K$ & $K+1$ & $K+1	$ 
		\\
		
		\rowcolor{yellow!10}
		$\omega_c$        & $\frac{2\pi}{M}$ & $\frac{K}{N}\frac{2\pi}{M}$ & $\frac{K}{N}\frac{2\pi}{M}$
		\\
		
		\rowcolor{yellow!10}
		$\epsilon_0$  & $2\cdot10^{-4}$ & $8\cdot10^{-5}$ & $2\cdot10^{-4}$
		\\
		
		\rowcolor{yellow!10}
		$u_0$  & $10^{-12}$ & $10^{-12}$ & $10^{-12}$
		\\
		
		\rowcolor{yellow!10}
		$\mathcal{K}$  & $\cbrac{0,1}$ & $\cbrac{0}$ & $\cbrac{0,1}$
		\\
		
		\rowcolor{yellow!10}
		$\delta$  & $2$ & $2$ & $2$
		\\
		
		\rowcolor{yellow!10}
		$f_i[k]$             & eq. \eqref{eq:fbmc_basis_slepian} & eq. \eqref{eq:fbmc_basis_cosine} & eq. \eqref{eq:fbmc_basis_cosine} 
		\\\hline
	\end{tabular}
	}
	\label{tab:fbmc_par}
\end{table}

Type-I configuration deploys $2K$ \ac{DPSS} with a passband of $\omega_s=2\pi/M$ to build the prototype filter, reasonably stringent interference tolerance $\epsilon_0$ and near null border samples. On the other hand, Type-II and Type-III filters are built through the summation $K+1$ cosines. Type-II imposes a higher reconstruction constraint, while Type-III focuses on a very fast spectrum decay. Furthermore, we set $\delta=2$ as it can easily make $\mb{Q}_{m,n}^{(3a)}$ and $\mb{Q}_{m,n}^{(3b)}$ positive semidefinite, according to eq. \eqref{eq:weyl1}. Concerning $\omega_c$, it is noteworthy mentioning that such a parameter is set around $2\pi/M$, given the subcarrier bandwidth and separation.  For Type-II and III, we set a more stringent $\omega_c$ to reduce the spectrum level within adjacent subcarriers.

The solution for the proposed design with the configurations described in Table \ref{tab:fbmc_par} takes two phases. First, the master problem \eqref{eq:prob_qcqp2a} is solved by using the Golden search, whereas the slave problem \eqref{eq:prob_qcqp2} was solved through MOSEK 8.0. For this class of problem, MOSEK casts the original \ac{QCQP} as a \ac{SOCP}, which is solved by the interior-point algorithm described in \cite{Andersen2003}. In order to achieve an easier implementation, \eqref{eq:prob_qcqp2a} was parsed {in}to MATLAB using the modeling language CVX. The solution of the proposed problem, \textit{i.e.}, the weights of the filters are provided at the Appendix, where some additional observations are offered.

\subsection{{Prototype Filter Performance Analysis}}
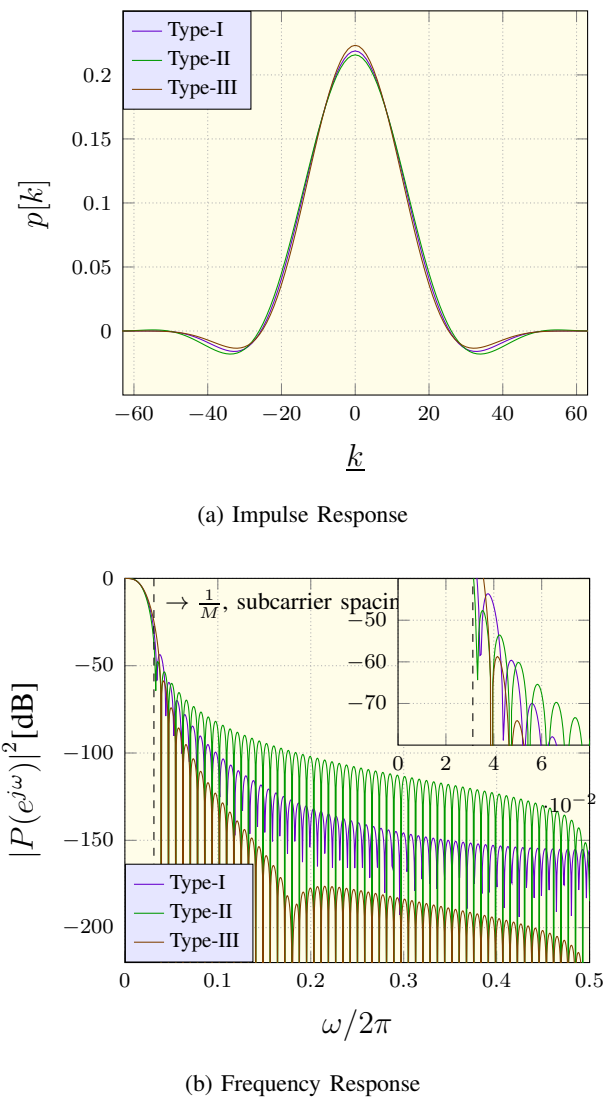
\begin{figure}[!b]
	\centering
	\vspace*{-9mm}
	\subfloat[Impulse Response\label{fbmc_pulse_time_prop_K=4}]
	{\begin{tikzpicture}[scale=1]
		\begin{axis}[ axisstyle,
					  legend columns=1,legend style={at={(0,1)},anchor=north west},
		              ylabel={$p[k]$}, xlabel={$\underline{k}$}, 
		              ytick={0,.05,.1,.15,.20},
		              xmin=-63,xmax=+63, ymin=-.05,ymax=.25,
		              mark repeat=4]
		    \legend{Type-I,Type-II,Type-III}
			\addplot[qcqp1, mark=none] table[x index=0,y index=1] {fbmc_pulse_time_qcqp1_K=4_M=32.dat};
			\addplot[qcqp2, mark=none] table[x index=0,y index=1] {fbmc_pulse_time_qcqp2_K=4_M=32.dat};
			\addplot[qcqp3, mark=none] table[x index=0,y index=1] {fbmc_pulse_time_qcqp3_K=4_M=32.dat};
		\end{axis}
	\end{tikzpicture}}
	\\
	\subfloat[Frequency Response\label{fbmc_pulse_freq_prop_K=4}]
	{\begin{tikzpicture}[scale=1]
		\begin{axis}[ axisstyle,
		              legend columns=1,legend style={at={(0,0)},anchor=south west},
		              ylabel={$\abs{P(e^{j\omega})}^2$[dB]}, xlabel={$\omega/2\pi$}, 
		              xmin=+0.0,xmax=+0.5,
		              ymin=-220,ymax=0]
		    \legend{Type-I,Type-II,Type-III}
			\addplot[qcqp1, mark=none] table[x index=0,y index=1] {fbmc_pulse_freq_qcqp1_K=4_M=32.dat};
   			\addplot[qcqp2, mark=none] table[x index=0,y index=1] {fbmc_pulse_freq_qcqp2_K=4_M=32.dat};
			\addplot[qcqp3, mark=none] table[x index=0,y index=1] {fbmc_pulse_freq_qcqp3_K=4_M=32.dat};		
			\addplot[color=black,dashed] coordinates{(1/32,0) (1/32,-300)};
			\node[anchor=north west] at (axis cs: 0.03125,0) {\footnotesize $\rightarrow\frac{1}{M}$, subcarrier spacing};
			\coordinate (posx) at (rel axis cs:1,1);
		\end{axis}
		\begin{axis}[ axisstyle,at={(posx)},anchor={north east},
		              width  = .25\textwidth, 
		              height = .23\textwidth,
    				  xtick={0,.02,.04,.06},
    				  ytick={-50,-60,-70},
    	              xmin=+0.00,xmax=+.08,
    	              ymin=-80,ymax=-40,
    	              ]
			\addplot[qcqp1, mark=none] table[x index=0,y index=1] {fbmc_pulse_freq_qcqp1_K=4_M=32.dat};
   			\addplot[qcqp2, mark=none] table[x index=0,y index=1] {fbmc_pulse_freq_qcqp2_K=4_M=32.dat};
			\addplot[qcqp3, mark=none] table[x index=0,y index=1] {fbmc_pulse_freq_qcqp3_K=4_M=32.dat};	
			\addplot[color=black,dashed] coordinates{(1/32,0) (1/32,-300)};
    	\end{axis}
	\end{tikzpicture}}
	\caption{Type-I, II and III pulses for $K=4$, $M=32$ and $L_p=KM+1$}
	\label{fig:fbmc_pulse_prop_K=4_M=32}
\end{figure}

In Fig. \ref{fig:fbmc_pulse_prop_K=4_M=32}, one can observe {both the impulse and frequency responses of} the prototype filters obtained via the proposed problem. As for the impulse response, all prototype filters are very similar. Nevertheless, the frequency response of Type-I, Type-II and Type-III are very distinguishable, besides presenting small sidelobes and fast spectrum decay. In particular, Type-I and  Type-III presented a very expressive spectrum decay, compared with Type-II. But by promoting a comparison among Type-I, Type-II, Type-III, \ac{EGF}, \ac{OFDP}, and Martin prototype filters,  one concludes that the proposed pulses achieved a superior spectrum decay with small sidelobes. Indeed, one may also observe that Type-II and Martin filter are very similar, but the former presented smaller sidelobes.

Let us now proceed a more precise analysis by deploying the figures of merit discussed in Section \ref{sec:fbmc_merit}. In this sense, Table \ref{tab:fbmc_perf}\footnote{The best filter value for each parameter is highlighted in green, while the worst is marked in red The \ac{EGF} with $\alpha=1/2$ and  $\alpha=2$ were not marked in red or green due to their poor symbols reconstruction {\it versus} spectrum performance trade-off.} summarizes the figures of merit of the proposed prototype filters {\it versus} the \ac{EGF}, \ac{OFDP}, and Martin. Initially, one can observe that the \ac{EGF} design provides a poor spectrum and a variable \ac{SIR} performance tuned by $\alpha$. In this sense, high values of $\alpha$ leads to a high \ac{SIR} but a poorer spectrum, while the opposed holds true, where $\alpha=1$ is typically considered a good trade-off. Moreover, Martin Filter presents a solid performance, with a high \ac{SIR} and a good spectrum performance, \textit{i.e.}, small sidelobes, fast energy decay and low frequency spread. However, such filter resulted in the poorest Heisenberg factor. Despite outperforming the \ac{EGF} in terms of spectrum, the  \ac{OFDP} design performed poorly, including the \ac{OoB} energy. Concerning the proposed prototype filters, Type-I delivers a reasonably high \ac{SIR} and a very competitive spectrum performance. Interestingly, Type-II filter showed a slightly superior performance, in almost all aspects, than Martin prototype filter, making it a very attractive choice. {It is worth mentioning that Type-II filter spectrum superiority over the Martin  prototype filter comes by its slightly reduced passband $\omega_c=\frac{K}{M}\frac{2\pi}{M}$. Furthermore, Type-II filter deploys an extra tone ($N=K+1$) when compared with the Martin filter ($K$ tones), improving the symbol reconstruction.} Finally, Type-III presented a similar \ac{SIR} performance to Type-I and the smallest \ac{MSL}. However, the \ac{OoB} energy $\overline{E}(2\pi/M)$ is among the highest.
\begin{table}[!t]
	\centering
	\caption{Prototype filters comparison considering $K=4$, $M=32$ and $L_p=KM+1$. } 
	\setlength{\tabcolsep}{2pt}
	\renewcommand{\arraystretch}{1.5}
	\scalebox{.85}
	{
	\begin{tabular}{cccccccc}
		\hline
		\rowcolor{blue!10}
		\textbf{Prototype Filter} & SIR[dB] & MSL[dB] & $D_k$ & $D_\nu$ & $\xi$  & $\overline{E}\rbrac{\frac{2\pi}{M}}$[dB] & $\overline{E}\rbrac{\frac{4\pi}{M}}$[dB]
		\\\hline
		\rowcolor{red!15}
		EGF ($\alpha=1/2$)   & 
		{$33.73$} & $-58.21$ & {$8.964$} & {$0.0101$} & $0.878$ & $-33.95$ & $-48.81$
		\\
		\rowcolor{red!15}
		EGF ($\alpha=2$)   & 
		{$114.48$} & {$-21.38$} & {$5.163$} & {$0.0176$} & {$0.874$} & {$-12.46$} & {$-20.67$}
		\\
		
		\rowcolor{yellow!10}
		EGF ($\alpha=1$)   & 
		$60.49$ & \colr{$-33.80$ }& {$6.457$} & \colr{$0.0126$} & {$0.976$} & \colr{$-19.69$} & \colr{$-33.50$}
		\\
		\rowcolor{yellow!10}
		Martin             & 
		$65.23$ & $-39.86$ & \colr{$8.784$} & {$0.0102$} & \colr{$0.884$} & $-45.61$ & $-70.60$
		\\
		\rowcolor{yellow!10}
		OFDP               & 
		$59.86$ & $-38.33$ & $7.842$ & $0.0109$ & $0.933$ & $-35.45$ & $-62.29$
		\\
		\rowcolor{yellow!10}
		Type-I              & 
		$52.74$ & $-43.63$ & $8.230$ & $0.0106$ & $0.915$ & $-42.30$ & $-82.96$
		\\
		\rowcolor{yellow!10}
		Type-II              & 
		{$68.09$} & $-47.68$ & $8.568$ & $0.0103$ & $0.897$ & {$-50.09$} & $-72.93$
		\\
		\rowcolor{yellow!10}
		Type-III              & 
		\colr{$51.25$} & {$-58.73$} & $7.877$ & $0.0108$ & $0.935$ & $-35.20$ & {$-100.57$}
		\\\hline
	\end{tabular}
	}
	\label{tab:fbmc_perf}
\end{table}

\begin{figure}[!b]
	\centering
	\vspace*{2mm}
	\begin{tikzpicture}[scale=1]
		\begin{axis}[ axisstyle,
					  legend columns=3,legend style={at={(1,1)},anchor=north east},
		              ylabel={$\overline{E}(\omega)$[dB]}, xlabel={$\omega/2\pi$}, 
		              xmin=0,xmax=0.25,ymin=-180,ymax=0,
		              mark repeat=50]
		\legend{EGF ($\alpha=1$),Martin,OFDP,Type-I,Type-II,Type-III}
		\addplot[egf]    table[x index=0,y index=1, each nth point=2] {fbmc_oob_egf_K=4_M=32.dat};
		\addplot[martin] table[x index=0,y index=1, each nth point=2] {fbmc_oob_martin_K=4_M=32.dat};
		\addplot[ofdp]   table[x index=0,y index=1, each nth point=2] {fbmc_oob_ofdp_K=4_M=32.dat};
		\addplot[qcqp1]  table[x index=0,y index=1, each nth point=2] {fbmc_oob_qcqp1_K=4_M=32.dat};
		\addplot[qcqp2]  table[x index=0,y index=1, each nth point=2] {fbmc_oob_qcqp2_K=4_M=32.dat};
		\addplot[qcqp3]  table[x index=0,y index=1, each nth point=2] {fbmc_oob_qcqp3_K=4_M=32.dat};
		\addplot[color=black,dashed] coordinates{(1/32,-180) (1/32,0)};
		\node[anchor=south west] at (axis cs: 0.03125,-180) {\footnotesize $\rightarrow\frac{1}{M}$, subcarrier spacing};
	\end{axis}
	\end{tikzpicture}
	\caption{Pulse \ac{OoB} Energy considering $K=4$, $M=32$ and $L_p=KM+1$}
	\label{fig:fbmc_oob_K=4}
\end{figure}
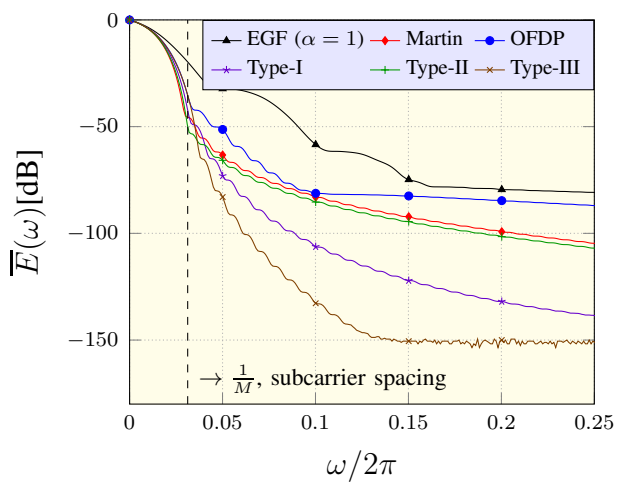

To complement the \ac{OoB} energy measurements presented in Table \ref{tab:fbmc_perf}, Fig. \ref{fig:fbmc_oob_K=4} portrays the \ac{OoB} energy for a broad range of frequencies. Through such figure, one can observe how fast the energy decays throughout the spectrum. Hence, faster decays indicate high spectrum efficiency and lower interference to adjacent bands.

\subsection{{Bit Error Rate Performance}}
As observed previously, the proposed prototype filters can offer a good performance in terms of \ac{SIR} and spectrum containment. However, we also present the performance in terms of {\ac{BER}} of an \ac{FBMC} system in a more contemporary application scenarios. In this sense, we chose a point-to-point MIMO-\ac{FBMC} system \cite{Rottenberg2017} deploying $N_t$ transmit antennas and $N_r$ receive antennas using \ac{V-BLAST} topology, \textit{i.e.}, spatial multiplexing mode. Furthermore, symbols are equalized via a \ac{ZF} MIMO equalizer with $L_w$ taps proposed in \cite{Ihalainen2011}, which  is more flexible than other methods such as \cite{Chen2017} as the equalizer does not require a flat channel response. The performance of the system was evaluated in a frequency selective Rayleigh channel considering three different scenarios described in Table \ref{tab:ber_setup}. Notice that the radio channel of Scenarios (A) and (C) are reasonably frequency selective, whereas Scenario (A) can be considered frequency flat per subcarrier.

\begin{figure}[!t]
	\begin{tikzpicture}[scale=1]
	\begin{semilogyaxis}[ axisstyle, grid=both,
legend style={at={(1,1)},anchor=north east},legend columns=3,
		                  xlabel={$E_b/N_0$[dB]}, ylabel={BER}, 
		                  xmin=00,xmax=60,ymin=3e-6,ymax=0.4,
		                  mark repeat=2]
		\legend{EGF ($\alpha=1$),Martin,OFDP,Type-I,Type-II,Type-III};
		\draw[->] (axis cs: 58,8E-5) |- (axis cs: 56,1.3E-3) node[align=right,anchor=east]{\scriptsize A};
		\addplot[egf,    mark phase=0] table[x index=0,y index=1] {fbmc_ber_egf_64qam_K=4_M=64_4x4_lw=7_eta=30.dat};
		\addplot[martin, mark phase=0] table[x index=0,y index=1] {fbmc_ber_martin_64qam_K=4_M=64_4x4_lw=7_eta=30.dat};
		\addplot[ofdp,   mark phase=0] table[x index=0,y index=1] {fbmc_ber_ofdp_64qam_K=4_M=64_4x4_lw=7_eta=30.dat};
		\addplot[qcqp1,  mark phase=0] table[x index=0,y index=1] {fbmc_ber_qcqp1_64qam_K=4_M=64_4x4_lw=7_eta=30.dat};
		\addplot[qcqp2,  mark phase=0] table[x index=0,y index=1] {fbmc_ber_qcqp2_64qam_K=4_M=64_4x4_lw=7_eta=30.dat};
		\addplot[qcqp3,  mark phase=0] table[x index=0,y index=1] {fbmc_ber_qcqp3_64qam_K=4_M=64_4x4_lw=7_eta=30.dat};
		\draw[->] (axis cs: 58,1.3E-5) |- (axis cs: 54,1.3E-5) node[align=right,anchor=east]{\scriptsize B};
		\addplot[egf,    mark phase=0] table[x index=0,y index=1] {fbmc_ber_egf_64qam_K=4_M=64_4x4_lw=1_eta=300.dat};
		\addplot[martin, mark phase=0] table[x index=0,y index=1] {fbmc_ber_martin_64qam_K=4_M=64_4x4_lw=1_eta=300.dat};
		\addplot[ofdp,   mark phase=0] table[x index=0,y index=1] {fbmc_ber_ofdp_64qam_K=4_M=64_4x4_lw=1_eta=300.dat};
		\addplot[qcqp1,  mark phase=0] table[x index=0,y index=1] {fbmc_ber_qcqp1_64qam_K=4_M=64_4x4_lw=1_eta=300.dat};
		\addplot[qcqp2,  mark phase=0] table[x index=0,y index=1] {fbmc_ber_qcqp2_64qam_K=4_M=64_4x4_lw=1_eta=300.dat};
		\addplot[qcqp3,  mark phase=0] table[x index=0,y index=1] {fbmc_ber_qcqp3_64qam_K=4_M=64_4x4_lw=1_eta=300.dat};
		\draw[->] (axis cs: 29,1.3E-5) |- (axis cs: 26,1.3E-5) node[align=right,anchor=east]{\scriptsize C};
		\addplot[egf,    mark phase=0] table[x index=0,y index=1] {fbmc_ber_egf_64qam_K=4_M=64_4x6_lw=1_eta=30.dat};
		\addplot[martin, mark phase=0] table[x index=0,y index=1] {fbmc_ber_martin_64qam_K=4_M=64_4x6_lw=1_eta=30.dat};
		\addplot[ofdp,   mark phase=0] table[x index=0,y index=1] {fbmc_ber_ofdp_64qam_K=4_M=64_4x6_lw=1_eta=30.dat};
		\addplot[qcqp1,  mark phase=0] table[x index=0,y index=1] {fbmc_ber_qcqp1_64qam_K=4_M=64_4x6_lw=1_eta=30.dat};
		\addplot[qcqp2,  mark phase=0] table[x index=0,y index=1] {fbmc_ber_qcqp2_64qam_K=4_M=64_4x6_lw=1_eta=30.dat};
		\addplot[qcqp3,  mark phase=0] table[x index=0,y index=1] {fbmc_ber_qcqp3_64qam_K=4_M=64_4x6_lw=1_eta=30.dat};	
		\addplot[egf,    mark phase=0] table[x index=0,y index=1] {fbmc_ber_egf12_64qam_K=4_M=64_4x4_lw=1_eta=300.dat};
		\addplot[egf,    mark phase=0] table[x index=0,y index=1] {fbmc_ber_egf12_64qam_K=4_M=64_4x6_lw=1_eta=30.dat};
	\end{semilogyaxis}
	\end{tikzpicture}
	\caption{\ac{BER} performance for different prototype filter choices.} 
	\label{fig:fbmc_ber_K=4}
\end{figure}
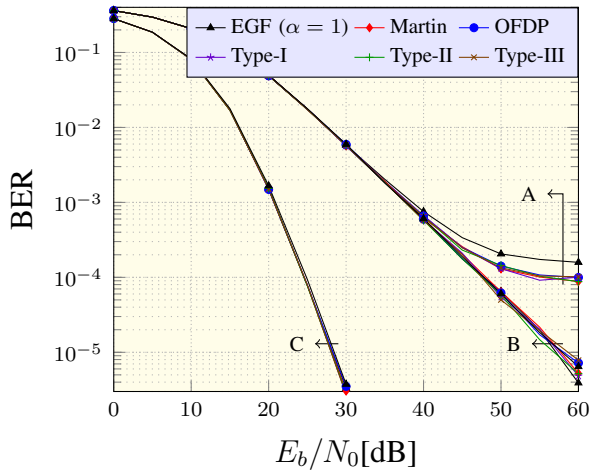

Fig. \ref{fig:fbmc_ber_K=4} depicts the \ac{BER} performance for all configurations and a wide normalized \ac{SNR} ($E_b/N_0$) range. Observing the results for Scenario (A), we can conclude that even a 7 taps equalization was unable to compensate the channel properly under a high \ac{SNR} regime, introducing a \ac{BER} floor. Yet in this configuration, we can also observe that the \ac{BER} floor level is proportional to $D_\nu$, where, for example, the \ac{EGF} filter performed poorly due to its higher frequency dispersion. For Scenario (B), all filters performed similarly in the analyzed $E_b/N_0$ range. Hence, despite differences in terms of the measured \ac{SIR}, the proposed prototype filters seems to be suitable in practical scenario. For Scenario (C), diversity improved the \ac{BER} considerably, making possible the usage of a single tap filter for channel equalization.
\begin{table}[!t]
	\centering
	\scriptsize
	\caption{Prototype filters comparison\label{tab:ber_setup} }
	\renewcommand{\arraystretch}{1.45}
	\begin{tabular}{ll}
		\hline
		\cellcolor{blue!10}                          & \cellcolor{yellow!10} MIMO V-BLAST,  $N_t\times N_r$ \cite{Rottenberg2017}\\[.5pt]
		\cellcolor{blue!10}                          & \cellcolor{yellow!10} FBMC multiplexing, No CP, $M=64$, $K=4$, $L_p=257$ \\[.5pt]
		\cellcolor{blue!10} 	                     & \cellcolor{yellow!10} Uncoded $8$-PAM, equivalent to $64$QAM in OFDM \\[.5pt]
		\multirow{-4}{*}{\cellcolor{blue!10}System}  & \cellcolor{yellow!10} $L_w$-taps Zero-Forcing Equalizer \cite{Ihalainen2011}\\[.5pt] \hline
		\multirow{-0}{*}{\cellcolor{blue!10}Channel}                          & \cellcolor{yellow!10} 
		\begin{tabular}{@{}l@{\hspace*{0mm}}l}
		Rayleigh \\
		
		\begin{tabular}{@{}ll}
		Exponential PDP \\\cite{Chayat1997}  
		\end{tabular}
		& \begin{tabular}{@{}l}
		$\expect{\abs{h[k]}^2} = \dfrac{1-e^{-1/\tau_\text{rms}}}{e^{-(1+10\tau_\text{rms})/\tau_\text{rms}}} e^{-\ell/\tau_\text{rms}}$\\
		$\ell=0,1,\cdots 10\tau_\text{rms}$
		\end{tabular}\\
		\begin{tabular}{@{}l}
		Coherence band\footnote{90th Coherence Band} \\\cite[eq. (4.31)]{Hampton2014}
		\end{tabular}
		 & $B_{c,90} = 1/(50\tau_\text{rms})$ \\
		Subchannel band & $B_{sc}=2/M$ \\[.5pt]
		Selectivity Index &$\eta_B = B_{c,90}/B_{sc}$ 
		\end{tabular} \\[.5pt]\hline
		\cellcolor{blue!10}                            & \cellcolor{yellow!10} (A) $4\times4$, $L_w=1$, $\eta_B=3$ \\[.5pt]
		\cellcolor{blue!10}                            & \cellcolor{yellow!10} (B) $4\times4$, $L_w=7$, $\eta_B=30$ \\[.5pt]
		\multirow{-3}{*}{\cellcolor{blue!10}Scenario}  & \cellcolor{yellow!10} (C) $4\times6$, $L_w=7$, $\eta_B=3$\\[.5pt]
		\hline
	\end{tabular}
	\vspace*{2mm}
\end{table}

At this point, it is noteworthy mentioning that the \ac{BER} floor can be established in two cases. The first case is the \ac{BER} floor generated due to the subcarrier frequency selectivity which took place in Scenario (A). The other \ac{BER} floor level is dictated by the self-interference of the prototype filter, which can be measured using the \ac{SIR} parameter. In Scenario (A), one could lower the \ac{BER} floor level by increasing the number of subcarriers or, conversely, by deploying a longer equalizer. However, such \ac{BER} floor never surpasses the one dictated by the self-interference of the prototype filter.

Despite possessing different \ac{SIR} levels, all the prototype filters presented similar \ac{BER} performance in each analyzed scenario, as depicted in Fig. \ref{fig:fbmc_ber_K=4}. Performance differences in Scenario (B) and (C) should arise only in higher \ac{SNR} level, which would require a much more demanding  computational resources. However, filters with extremely high \ac{SIR} levels may not be required as systems typically do not operate in a \ac{SNR} level around $70-100$[dB], where self-interference should take place. Despite Type-I and Type-III possessing lower \ac{SIR} levels, they do not introduce noticeable performance losses in the presented \ac{SNR} range, which covers most practical application scenarios. Thus, the keypoint of this analysis is to demonstrate that the proposed prototype filters impose no symbol reconstruction drawbacks, while offering a considerable spectrum improvement, making them an interesting choice for current and future applications.

\section{Conclusions and Final Remarks}\label{sec:conclusions}
Throughout this paper, we proposed a prototype filter design framework capable of providing high symbol reconstruction performance and desirable spectral features. The main difference between the proposed design methodology and other available options is its solution method based on convex optimization. In this sense, we were able to write such a complex design into a convex \ac{QCQP} problem guided by a line search, which benefits from powerful available optimization tools. As a result, we proposed three prototype filters. The Type-I and Type-III filters presented very fast spectrum decays, with the \acs{MSL} ranging from $-45$ to $-58$[dB] and reasonably high SIR values. On the other hand, the Type-II filter demonstrated to be slightly superior, in almost all aspects, then the Mirabbasi-Martin design, which is a standard choice for \ac{FBMC} systems. Thus, the proposed design methodology showed to be both flexible and effective, given the superior spectrum performance achieved by the proposed prototype filters.


\begin{thebibliography}{10}
\providecommand{\url}[1]{#1}
\csname url@samestyle\endcsname
\providecommand{\newblock}{\relax}
\providecommand{\bibinfo}[2]{#2}
\providecommand{\BIBentrySTDinterwordspacing}{\spaceskip=0pt\relax}
\providecommand{\BIBentryALTinterwordstretchfactor}{4}
\providecommand{\BIBentryALTinterwordspacing}{\spaceskip=\fontdimen2\font plus
\BIBentryALTinterwordstretchfactor\fontdimen3\font minus
  \fontdimen4\font\relax}
\providecommand{\BIBforeignlanguage}[2]{{%
\expandafter\ifx\csname l@#1\endcsname\relax
\typeout{** WARNING: IEEEtran.bst: No hyphenation pattern has been}%
\typeout{** loaded for the language `#1'. Using the pattern for}%
\typeout{** the default language instead.}%
\else
\language=\csname l@#1\endcsname
\fi
#2}}
\providecommand{\BIBdecl}{\relax}
\BIBdecl

\bibitem{Hwang2009}
T.~Hwang, C.~Yang, G.~Wu, S.~Li, and G.~Y. Li, ``{OFDM and Its Wireless
  Applications: A Survey},'' \emph{IEEE Transactions on Vehicular Technology},
  vol.~58, no.~4, pp. 1673--1694, May 2009.

\bibitem{Freeman2016}
H.~Freeman and A.~Dutta, ``{5G Perspective [The President's Page]},''
  \emph{IEEE Communications Magazine}, vol.~54, no.~5, pp. 4--5, May 2016.

\bibitem{Wunder2014}
G.~Wunder, P.~Jung, M.~Kasparick, T.~Wild, F.~Schaich, Y.~Chen, S.~T. Brink,
  I.~Gaspar, N.~Michailow, A.~Festag, L.~Mendes, N.~Cassiau, D.~Ktenas,
  M.~Dryjanski, S.~Pietrzyk, B.~Eged, P.~Vago, and F.~Wiedmann, ``{5GNOW:
  Non-Orthogonal, Asynchronous Waveforms for Future Mobile Applications},''
  \emph{IEEE Communications Magazine}, vol.~52, no.~2, pp. 97--105, February
  2014.

\bibitem{Cassiau2013}
N.~Cassiau, D.~Ktenas, and J.~B. Dore, ``{Time and Frequency Synchronization
  for CoMP with FBMC},'' in \emph{ISWCS 2013; The Tenth International Symposium
  on Wireless Communication Systems}, Aug 2013, pp. 1--5.

\bibitem{Farhang2014}
A.~Farhang, N.~Marchetti, F.~Figueiredo, and J.~P. Miranda, ``{Massive MIMO and
  Waveform Design for 5th Generation Wireless Communication Systems},'' in
  \emph{1st International Conference on 5G for Ubiquitous Connectivity}, Nov
  2014, pp. 70--75.

\bibitem{Boccardi2014}
F.~Boccardi, R.~W. Heath, A.~Lozano, T.~L. Marzetta, and P.~Popovski, ``{Five
  Disruptive Technology Directions for 5G},'' \emph{IEEE Communications
  Magazine}, vol.~52, no.~2, pp. 74--80, February 2014.

\bibitem{Metis}
M.~Schellmann, ``{Mobile and Wireless Communications Enablers for the
  Twenty-twenty Information Society (METIS)},'' Tech. Rep.
  ICT-317669-METIS/D2.4 Proposed, Feb 2015.

\bibitem{Phydias}
A.~Viholainen, M.~Bellanger, and M.~Huchard, ``{PHYDYAS 007 - PHYsical layer
  for DYnamic AccesS and Cognitive Radio},'' Tech. Rep. ICT-211887, Jan 2009.

\bibitem{Fang2013}
J.~Fang, Z.~You, I.~T. Lu, J.~Li, and R.~Yang, ``{Comparisons of Filter Bank
  Multicarrier Systems},'' in \emph{{2013 IEEE Long Island Systems,
  Applications and Technology Conference (LISAT)}}.\hskip 1em plus 0.5em minus
  0.4em\relax New York, USA: {IEEE}, 2013, pp. 1--6.

\bibitem{Waldhauser2006}
D.~S. Waldhauser, L.~G. Baltar, and J.~A. Nossek, ``{Comparison of Filter Bank
  Based Multicarrier Systems with OFDM},'' in \emph{{Apccas 2006 - 2006 IEEE
  Asia Pacific Conference on Circuits and Systems}}, 2006, pp. 976--979.

\bibitem{Farhang2011}
B.~Farhang-Boroujeny, ``{OFDM Versus Filter Bank Multicarrier},'' \emph{IEEE
  Signal Processing Magazine}, vol.~28, no.~3, pp. 92--112, May 2011.

\bibitem{Lele2008b}
C.~Lele, R.~Legouable, and P.~Siohan, ``{Channel Estimation with Scattered
  Pilots in OFDM/OQAM},'' in \emph{2008 IEEE 9th Workshop on Signal Processing
  Advances in Wireless Communications}, July 2008, pp. 286--290.

\bibitem{Cui2016}
W.~Cui, D.~Qu, T.~Jiang, and B.~Farhang-Boroujeny, ``{Coded Auxiliary Pilots
  for Channel Estimation in FBMC-OQAM Systems},'' \emph{IEEE Transactions on
  Vehicular Technology}, vol.~65, no.~5, pp. 2936--2946, May 2016.

\bibitem{Rahmatallah2013}
Y.~Rahmatallah and S.~Mohan, ``{Peak-To-Average Power Ratio Reduction in OFDM
  Systems: A Survey And Taxonomy},'' \emph{Communications Surveys {\&}
  Tutorials, IEEE}, vol.~15, no.~4, pp. 1567--1592, 2013.

\bibitem{Bulusu2015}
S.~S. K.~C. Bulusu, H.~Shaiek, and D.~Roviras, ``{Reduction of PAPR of
  FBMC-OQAM Systems by Dispersive Tone Reservation Technique},'' in \emph{2015
  International Symposium on Wireless Communication Systems (ISWCS)}, Aug 2015,
  pp. 561--565.

\bibitem{Kumar2016}
R.~Kumar and A.~Tyagi, ``{Computationally Efficient Mask-Compliant Spectral
  Precoder for OFDM Cognitive Radio},'' \emph{IEEE Transactions on Cognitive
  Communications and Networking}, vol.~2, no.~1, pp. 15--23, March 2016.

\bibitem{Zhang2017}
L.~Zhang, P.~Xiao, A.~Zafar, A.~u.~Quddus, and R.~Tafazolli, ``{FBMC System: An
  Insight Into Doubly Dispersive Channel Impact},'' \emph{IEEE Transactions on
  Vehicular Technology}, vol.~66, no.~5, pp. 3942--3956, May 2017.

\bibitem{Siohan2002}
P.~Siohan, C.~Siclet, and N.~Lacaille, ``{Analysis and Design of OFDM/OQAM
  Systems Based on Filterbank Theory},'' \emph{IEEE Transactions on Signal
  Processing}, vol.~50, no.~5, pp. 1170--1183, 2002.

\bibitem{Mirabbasi2003}
S.~Mirabbasi and K.~Martin, ``{Overlapped Complex-Modulated Transmultiplexer
  Filters with Simplified Design and Superior Stopbands},'' \emph{IEEE
  Transactions on Circuits and Systems II: Analog and Digital Signal
  Processing}, vol.~50, no.~8, pp. 456--469, Aug 2003.

\bibitem{Bellanger2001}
M.~G. Bellanger, ``{Specification and Design of a Prototype Filter for Filter
  Bank Based Multicarrier Transmission},'' in \emph{2001 IEEE International
  Conference on Acoustics, Speech, and Signal Processing. Proceedings (Cat.
  No.01CH37221)}, vol.~4, 2001, pp. 2417--2420 vol.4.

\bibitem{Vaydianathan1993}
P.~P. Vaidyanathan, \emph{{Multirate Systems and Filter Banks}}.\hskip 1em plus
  0.5em minus 0.4em\relax Upper Saddle River, NJ, USA: Prentice-Hall, Inc.,
  1993.

\bibitem{Haas1997}
R.~Haas and J.-C. Belfiore, ``{A Time-Frequency Well-localized Pulse for
  Multiple Carrier Transmission},'' \emph{Wireless Personal Communications},
  vol.~5, no.~1, pp. 1--18, Jul 1997.

\bibitem{Prakash2013}
J.~A. Prakash and G.~R. Reddy, ``{Efficient Prototype Filter Design for Filter
  Bank Multicarrier (FBMC) System Based on Ambiguity Function Analysis of
  Hermite polynomials},'' in \emph{2013 International Mutli-Conference on
  Automation, Computing, Communication, Control and Compressed Sensing
  (iMac4s)}, March 2013, pp. 580--585.

\bibitem{Aminjavaheri2017}
A.~Aminjavaheri, A.~Farhang, L.~E. Doyle, and B.~Farhang-Boroujeny,
  ``{Prototype Filter Design for FBMC in Massive MIMO Channels},'' in
  \emph{2017 IEEE International Conference on Communications (ICC)}, May 2017,
  pp. 1--6.

\bibitem{Afif2016}
M.~Dohler and T.~Nakamura, \emph{{5G Mobile and Wireless Communications
  Technology}}, A.~Osseiran, J.~F. Monserrat, and P.~Marsch, Eds.\hskip 1em
  plus 0.5em minus 0.4em\relax Cambridge University Press, 2016.

\bibitem{Sahin2014}
A.~Sahin, I.~Guvenc, and H.~Arslan, ``{A Survey on Multicarrier Communications:
  Prototype Filters, Lattice Structures, and Implementation Aspects},''
  \emph{IEEE Communications Surveys Tutorials}, vol.~16, no.~3, pp. 1312--1338,
  Third 2014.

\bibitem{Lefloch1995}
B.~L. Floch, M.~Alard, and C.~Berrou, ``{Coded Orthogonal Frequency Dvision
  Multiplex [TV Broadcasting]},'' \emph{Proceedings of the IEEE}, vol.~83,
  no.~6, pp. 982--996, Jun 1995.

\bibitem{Siohan2000}
P.~Siohan and C.~Roche, ``{Cosine-Modulated Filterbanks Based on Extended
  Gaussian Functions},'' \emph{IEEE Transactions on Signal Processing},
  vol.~48, no.~11, pp. 3052--3061, 2000.

\bibitem{Horn1985}
R.~A. Horn and C.~R. Johnson, \emph{{Matrix Analysis}}.\hskip 1em plus 0.5em
  minus 0.4em\relax New York, USA: Cambridge University Press, 1985.

\bibitem{Slepian1978}
D.~Slepian, ``{Prolate Spheroidal Wave Functions, Fourier analysis, and
  Uncertainty V: The Discrete Case},'' \emph{The Bell System Technical
  Journal}, vol.~57, no.~5, pp. 1371--1430, May 1978.

\bibitem{Moore2004}
I.~C.Moore and M.~Cada, ``Prolate spheroidal wave functions, fourier analysis,
  and uncertainty v: the discrete case,'' \emph{{Prolate Spheroidal Wave
  functions, an Introduction to the Slepian Series and its Properties}},
  vol.~16, no.~3, pp. 208--230, Mar 2004.

\bibitem{Vahlin1994}
A.~Vahlin and N.~Holte, ``{Optimal Finite Duration Pulses for OFDM},'' in
  \emph{Global Telecommunications Conference, 1994}, 1994, pp. 258--262.

\bibitem{Lu2011}
C.~Lu, S.-C. Fang, Q.~Jin, Z.~Wang, and W.~Xing, ``{KKT Solution and Conic
  Relaxation for Solving Quadratically Constrained Quadratic Programming
  Problems},'' \emph{SIAM Journal on Optimization}, vol.~21, no.~4, pp.
  1475--1490, Dec 2011.

\bibitem{Boyd2004}
S.~Boyd and L.~Vandenberghe, \emph{{Convex Optimization}}.\hskip 1em plus 0.5em
  minus 0.4em\relax New York, NY, USA: Cambridge University Press, 2004.

\bibitem{Luo2010}
Z.~q.~Luo, W.~k.~Ma, A.~M. c.~So, Y.~Ye, and S.~Zhang, ``{Semidefinite
  Relaxation of Quadratic Optimization Problems},'' \emph{IEEE Signal
  Processing Magazine}, vol.~27, no.~3, pp. 20--34, May 2010.

\bibitem{Andersen2003}
E.~Andersen, C.~Roos, and T.~Terlaky, ``{On implementing a primal-dual
  interior-point method for conic quadratic optimization},'' \emph{Mathematical
  Programming}, vol.~95, no.~2, pp. 249--277, Feb 2003.

\bibitem{Rottenberg2017}
F.~Rottenberg, X.~Mestre, F.~Horlin, and J.~Louveaux, ``{Single-Tap Precoders
  and Decoders for Multiuser MIMO FBMC-OQAM Under Strong Channel Frequency
  Selectivity},'' \emph{IEEE Transactions on Signal Processing}, vol.~65,
  no.~3, pp. 587--600, Feb 2017.

\bibitem{Ihalainen2011}
T.~Ihalainen, A.~Ikhlef, J.~Louveaux, and M.~Renfors, ``{Channel Equalization
  for Multi-Antenna FBMC/OQAM Receivers},'' \emph{IEEE Transactions on
  Vehicular Technology}, vol.~60, no.~5, pp. 2070--2085, Jun 2011.

\bibitem{Chen2017}
C.~W. Chen and F.~Maehara, ``{An Enhanced MMSE Subchannel Decision Feedback
  Equalizer with ICI Suppression for FBMC/OQAM Systems},'' in \emph{2017
  International Conference on Computing, Networking and Communications (ICNC)},
  Jan 2017, pp. 1041--1045.

\bibitem{Chayat1997}
N.~Chayat, ``{Tentative Criteria for Comparison of Modulation Methods},'' IEEE,
  Tech. Rep., 1997.

\bibitem{Hampton2014}
J.~R. Hampton, \emph{{Introduction to MIMO Communications}}, 1st~ed.\hskip 1em
  plus 0.5em minus 0.4em\relax New York, USA: Cambridge University Press, 2014.

\end{thebibliography}


\appendix
\section*{Prototype Filter Weights}
The weights of the proposed prototype filter are presented in Table \ref{tab:fbmc_weights}.
\begin{table}[!htbp]
	\centering
	\caption{Filter weights for Type-I, Type-II and Type-III configurations with $K=4$, $M=32$, $L_p=KM+1$}
	\scalebox{.85}
	{
	\setlength{\tabcolsep}{3pt}
	\begin{tabular}{cccc}
		\hline
		\rowcolor{blue!10}
		$c_i$ & \textbf{Type-I} & \textbf{Type-II} & \textbf{Type-III} \\\hline
		\rowcolor{yellow!10}
		$c_{0}$ & $9.179317816790\cdot10^{-1}$ & $5.016511380872\cdot10^{-1}$ & $4.993086025524\cdot10^{-1}$ \\
		\rowcolor{yellow!10}
		$c_{1}$ & $3.802162534407\cdot10^{-1}$ & $6.897038048179\cdot10^{-1}$ & $6.777473126670\cdot10^{-1}$ \\
		\rowcolor{yellow!10}
		$c_{2}$ & $1.077526750194\cdot10^{-1}$ & $5.039449735142\cdot10^{-1}$ & $5.037266848356\cdot10^{-1}$ \\
		\rowcolor{yellow!10}
		$c_{3}$ & $2.456277538185\cdot10^{-2}$ & $1.795258480584\cdot10^{-1}$ & $2.213401597940\cdot10^{-1}$ \\
		\rowcolor{yellow!10}
		$c_{4}$ & $4.639914990515\cdot10^{-3}$ & $9.191524770412\cdot10^{-3}$ & $4.093046350246\cdot10^{-2}$ \\
		\hline
		\rowcolor{yellow!10}
		$c_{5}$ & $1.306778847145\cdot10^{-3}$ & $-$ & $-$ \\
		\rowcolor{yellow!10}
		$c_{6}$ & $1.577770437750\cdot10^{-3}$ & $-$ & $-$ \\
		\rowcolor{yellow!10}
		$c_{7}$ & $3.721905313771\cdot10^{-4}$ & $-$ & $-$ \\			
		\hline
	\end{tabular}
	}
	\label{tab:fbmc_weights}
\end{table}

In particular, if $\mb{F}$ is taken as a cosine basis (Type-II and Type-III) scaling the number of subcarriers for a given overlapping factor $K$ is an easier task due to the frequency sampling feature of such basis. In this sense, even if the coefficients provided in Table \ref{tab:fbmc_weights} are derived for a specific value of $M$, one can still scale the prototype filter for larger values of $M$. This can be achieve by correctly scaling the frequency components via
\begin{equation}
	p[k] = \displaystyle\sum_{i=0}^{N-1} c_i' \cos\rbrac{ \dfrac{2\pi}{KM}i\underline{k} },
\end{equation}
where
\begin{equation}
c_i' = 
\left\lbrace
\begin{array}{ll}
1, & i = 0\\
\sqrt{\dfrac{2L_p}{L_p+1}} \dfrac{c_i}{c_0}, & \text{otherwise}
\end{array}
\right. .
\end{equation}

\section*{Acknowledgments}
This work was supported in part by the National Council for Scientific and Technological Development (CNPq) of Brazil under grants 404079/2016-4 and 304066/2015-0, in part by CAPES-Brazil (PhD scholarship), and in part by Londrina State University, Parana State Government (UEL).

\newpage
\begin{IEEEbiography}[{\includegraphics[width=1in,height=1.25in,clip,keepaspectratio]{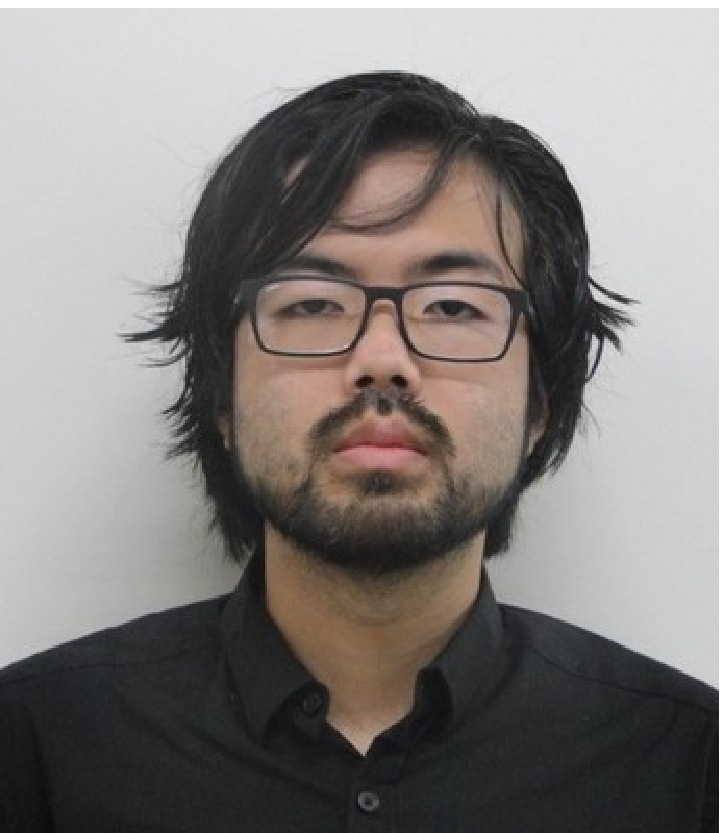}}]
{RICARDO TADASHI KOBAYASHI} Received the B.Tech. and Master degree, both in Electrical Engineering from the State University of Londrina, Londrina, Brazil in 2014 and 2016, respectively. Currently, he is with State University of Londrina as a P.hD. student working toward his Doctorate EE degree. His research interests lie in communications and signal processing, including MIMO detection techniques, equalization for multicarrier systems, optimization aspects of communications, convex optimization, cognitive radio techniques and filter design for multicarrier applications.
\end{IEEEbiography}

\vspace{-40mm}
\begin{IEEEbiography}[{\includegraphics[width=1in,height=1.25in,clip,keepaspectratio]{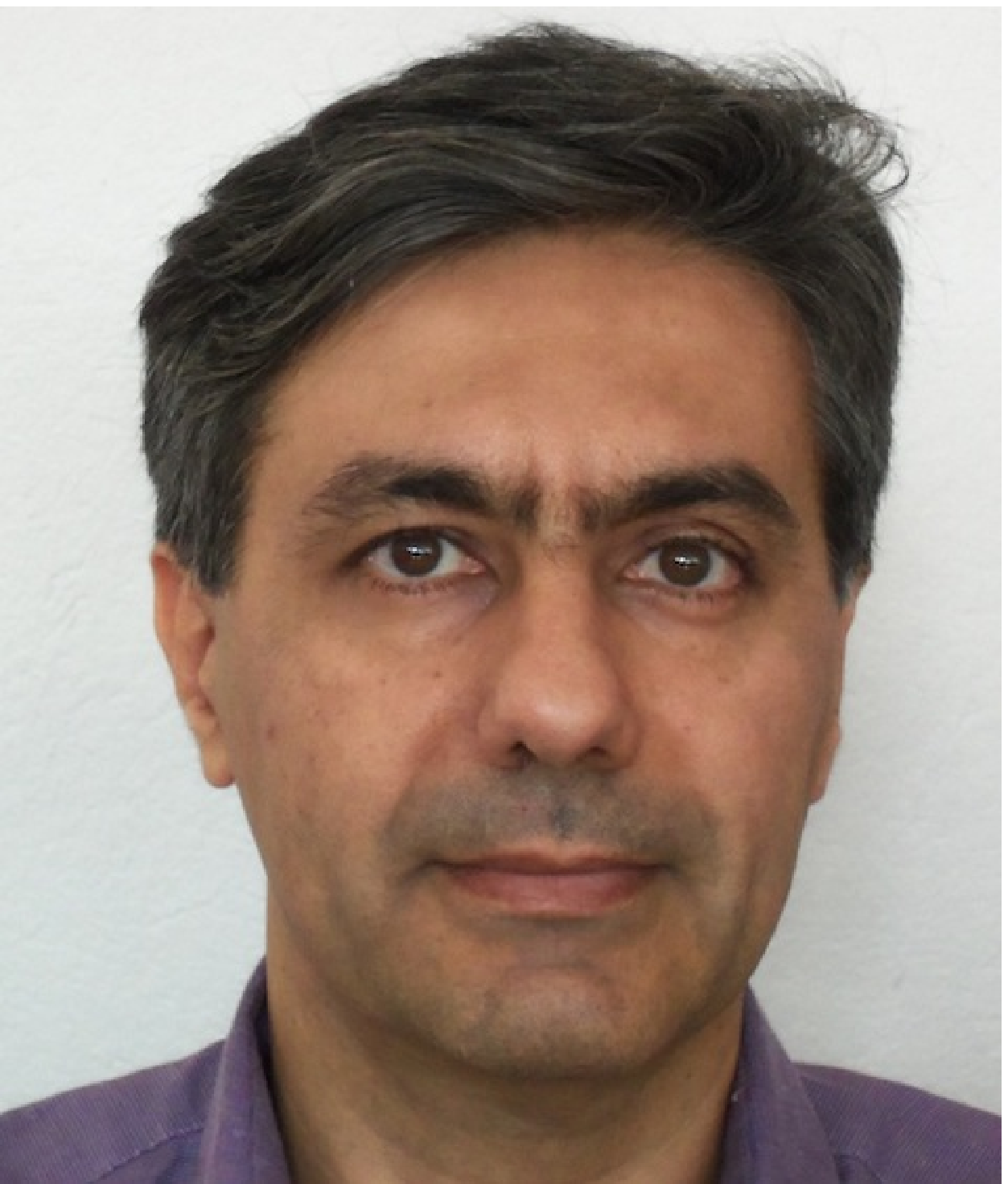}}]
{TAUFIK ABRÃO} (IEEE-SM'12, SM-SBrT) received the B.S., M.Sc., and Ph.D. degrees in electrical engineering from the Polytechnic School of the University of São Paulo, Brazil, in 1992, 1996, and 2001, respectively. Since March 1997, he has been with the Communications Group, Department of Electrical Engineering, Londrina State University, Londrina, Brazil, where he is currently an Associate Professor of Communications Engineering and the head of the Telecommunication and Signal Processing Group. He has been a Guest Researcher in the Connectivity Group at Aalborg University, DK (July-Oct. 2018). In 2012, he was an Academic Visitor with the Communications, Signal Processing, and Control Research Group, University of Southampton, Southampton, U.K. From 2007 to 2008 he was a Postdoctoral Researcher with the Department of Signal Theory and Communications, Polytechnic University of Catalonia (TSC/UPC), Barcelona, Spain. He has participated in several projects funded by government agencies and industrial companies. He has supervised 24 M.Sc.,  four Ph.D. students, and two postdocs. He has co-authored 11 book chapters on mobile radio communications He is involved in editorial board activities of six journals in the wireless communication area, and he has served as TCP member in several symposium and conferences. He has been served as an Editor for the IEEE Communications Surveys $\&$ Tutorials since 2013, IET Journal of Engineering since 2014, and IEEE Access since 2016 and Transactions on Emerging Telecommunications Technologies (ETT-Wiley) since 2018. He is a senior member of IEEE and SBrT-Brazil. His current research interests include communications and signal processing, especially in massive MIMO multiuser detection and estimation, ultra-reliable low latency communications (URLLC), machine-type communication (MTC), resource allocation, as well as heuristic and convex optimization aspects of 4G and 5G wireless systems. He has co-authored +240 research papers published in specialized/international journals and conferences.
\end{IEEEbiography}

\end{document}